\documentclass[twocolumn]{aastex701}

\usepackage{comment}
\usepackage{color}
\usepackage{graphicx}
\usepackage{natbib}
\usepackage{sidecap}
\usepackage[many]{tcolorbox}
\usepackage{wrapfig}
\usepackage{subfigure}

\newcommand{\HI}{\ion{H}{1}}
\newcommand{\DI}{\ion{D}{1}}
\newcommand{\HeI}{\ion{He}{1}}
\newcommand{\HeII}{\ion{He}{2}}
\newcommand{\MgII}{\ion{Mg}{2}}
\newcommand{\FeII}{\ion{Fe}{2}}

\newcommand{\NaI}{\ion{Na}{1}}
\newcommand{\Lya}{Ly$\alpha$}
\newcommand{\kms}{km s$^{-1}$}
\newcommand{\NHI}{N(HI)}
\newcommand{\logNHI}{log$_{10}$[\NHI/cm$^{-2}$]}
\newcommand{\nhiavg}{$\bar{n}$(HI)}
\newcommand{\nhitrue}{$n$(HI)}

\newcommand{\limitingdistance}{100}
\newcommand{\ntargets}{164}
\newcommand{\lowerresidual}{0.20}
\newcommand{\upperresidual}{0.48}

\begin{document}

\title{Toward a 2D \HI\ Map of the Local Interstellar Medium}

\author[0000-0002-1176-3391]{Allison Youngblood}
\affiliation{Exoplanets and Stellar Astrophysics Laboratory, NASA Goddard Space Flight Center, Greenbelt, MD 20771, USA}
\email{allison.a.youngblood@nasa.gov}

\author[0000-0002-1002-3674]{Kevin France}
\affiliation{University of Colorado, 600 UCB, Boulder, CO 80309, USA}
\email{kevin.france@colorado.edu}

\author[0000-0003-3071-8358]{Tommi Koskinen}
\affiliation{Lunar and Planetary Laboratory, University of Arizona, Tucson, AZ 85721, USA}
\email{tommi@lpl.arizona.edu}

\author[0000-0002-3783-5509]{James Paul Mason}
\affiliation{Johns Hopkins University Applied Physics Laboratory, 11000 Johns Hopkins Rd, Laurel, MD 20723, USA}
\email{james.mason@jhuapl.edu}

\author[0000-0003-3786-3486]{Seth Redfield}
\affiliation{Astronomy Department and Van Vleck Observatory, Wesleyan University, Middletown, CT 06459, USA}
\email{sredfield@wesleyan.edu}

\author[0000-0002-4998-0893]{Brian E. Wood}
\affiliation{Naval Research Laboratory, Space Science Division, Washington, DC 20375, USA}
\email{brian.e.wood26.civ@us.navy.mil}

\author[0000-0002-9148-034X]{Vincent Bourrier}
\affiliation{Observatoire Astronomique de l'Universit\'e de Gen\`eve, Chemin Pegasi 51b, CH-1290 Versoix, Switzerland}
\email{vincent.bourrier@unige.ch}

\author[0000-0002-2248-3838]{Leonardo dos Santos}
\affiliation{Space Telescope Science Institute, 3700 San Martin Drive, Baltimore, MD 21218, USA}
\email{ldsantos@stsci.edu}

\author[0000-0002-8828-6386]{Christopher Johns-Krull}
\affiliation{Department of Physics and Astronomy, Rice University, 6100 Main Street, Houston, TX 77005, USA}
\email{cmj@rice.edu}

\author[0000-0002-3641-6636]{George W. King}
\affiliation{Department of Astronomy, University of Michigan, Ann Arbor, MI 48109, USA}
\email{kinggw@umich.edu}

\author[0000-0003-4446-3181]{Jeffrey L. Linsky}
\affiliation{JILA, University of Colorado and NIST, Boulder, CO, 80309-0440, USA}
\email{jlinsky@jila.colorado.edu}

\author[0000-0002-1046-025X]{Sarah Peacock}
\affiliation{University of Maryland, Baltimore County, Baltimore, MD 21250, USA}
\affiliation{Exoplanets and Stellar Astrophysics Laboratory, NASA Goddard Space Flight Center, Greenbelt, MD 20771, USA}
\email{sarah.r.peacock@nasa.gov}

\begin{abstract}
Obtaining a complete census of gas in the local interstellar medium ($<$100 pc) is challenging given the limited available tracers of the warm, partially-ionized medium. Medium-to-high resolution UV absorption spectroscopy toward individual nearby stars is the primary method used, and incomplete spatial sampling of this complex medium makes a global map of the material difficult. Using \HI\ column density measurements derived from \HI\ \Lya\ spectroscopy toward \ntargets\ stars inside \limitingdistance\ pc, we have generated 2D spatially-interpolated \NHI\ maps for different distance shells. Based on the area-weighted sky averages, we find that sightlines inside 10 pc typically have \logNHI\ $\sim$ {17.9}. For greater distance shells, \logNHI\ increases to {18.3} (10-20 pc), then to 18.4 (20-70 pc), and finally increasing to 18.6 (70-100 pc). This last increase is likely associated with the detection of the Local Bubble boundary, thus making the plateau of column density from 20-70 pc notable and suggestive of the rarity of warm LISM material beyond $\sim$10-20 pc. We estimate that the uncertainties associated with \NHI\ values inferred from the interpolated sky maps are approximately inversely correlated with the number of samples in each distance shell and are in the range of \lowerresidual-\upperresidual\ dex, compared to 0.01-0.30 dex {typically determined} from direct \Lya\ observations. We discuss the impact of these uncertainties on ISM corrections of EUV and \Lya\ observations for nearby stars. Denser spatial sampling of the sky via UV absorption spectroscopy of nearby stars is required to improve the accuracy of these \NHI\ estimates.
\end{abstract}

\section{Introduction} \label{sec:intro}

The Sun resides within the Local Cavity (also known as The Local Bubble), a region of low-density material with little dust extending to $\sim$100 pc in every direction \citep{Welsh2010,Lallement2022,Zucker2023}. Various photoionization sources are scattered unevenly through the Local Cavity, creating structured clouds of different ionization fractions. Large column densities derived from \HI\ 21 cm maps indicated that the interstellar medium (ISM) would be completely opaque to radiation $<$912 \AA\ \citep{Aller1959}, and it was a surprise to discover that within $\sim$100 pc of the solar system, the local ISM (LISM) has gas volume densities an order of magnitude lower than much of the gas probed by 21 cm radiation \citep{Cruddace1974,McClintock1975,Cox1987}.

\begin{figure*}
    \centering
    \includegraphics[width=\textwidth]{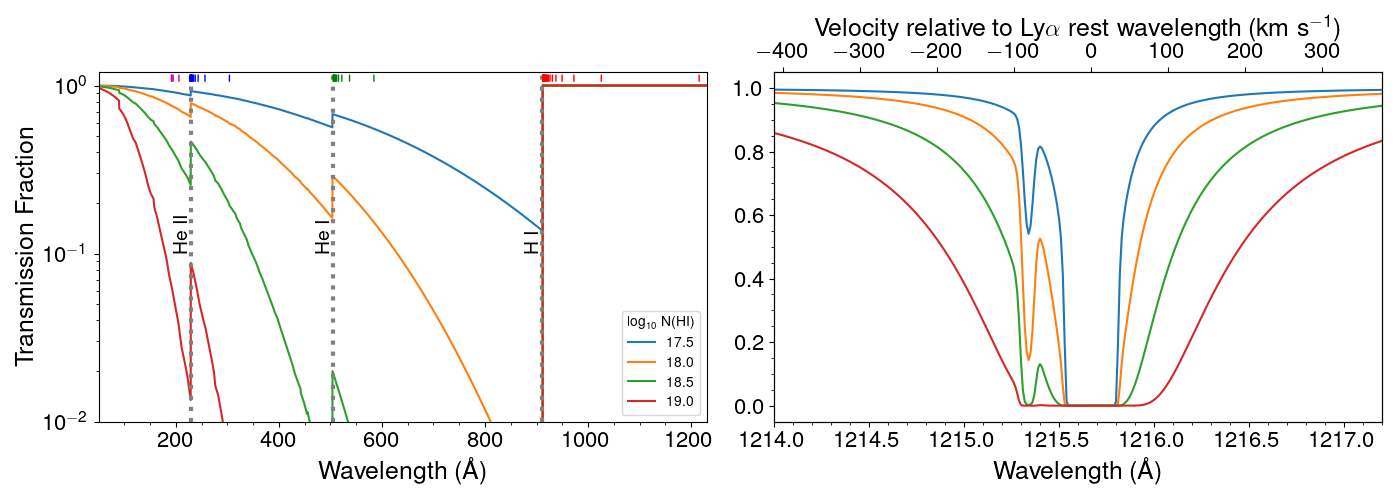}
    \caption{Left: The transmission fraction ($e^{-\tau}$) of \HI, \HeI, and \HeII\ gas in the ISM as a function of wavelength from 50-1230 \AA\ for a range of \HI\ column densities typical for the LISM. Only bound-free (continuum) opacity is shown. N(\HeI)/\NHI=0.08 and N(\HeII)/N(\HeI)=1.5 (upper limit) were assumed \citep{Dupuis1995}. The ionization edges of \HeII\ (229 \AA), \HeI\ (504 \AA), and \HI\ (912 \AA) are marked with vertical lines. At $\lambda>$912 \AA, transmission is unity. Colored vertical ticks mark the wavelengths of the ground state series of \HI\ (red), \HeI\ (green), and \HeII\ (blue), along with \HeI\ autoionization lines (magenta), where additional opacity is expected, but not shown. Right: A close-up of the transmission curve of \HI\ and \DI\ at \Lya\ (1215.67 and 1215.34 \AA, respectively) as a function of wavelength and velocity for the same range of \NHI\ values as the left panel. N(\DI)/N(\HI) = 1.56$\times$10$^{-5}$, $b_{HI}$ = 10.8 \kms, and high spectral resolution (R$>$100,000) were assumed. }
    \label{fig:attenuation_curve}
\end{figure*}

The LISM's partially ionized gas strongly absorbs H and He resonance lines and bound-free transitions, as well as low-ionization metal species, all of which lie in the ultraviolet (UV; 100-3000 \AA). Extreme-UV (EUV) photons (100-912 \AA) are strongly absorbed by bound-free transitions of H, He, and He$^{+}$ \citep{Rumph1994}, each of which have their ionization edges at 912, 504, and 228 \AA, respectively (Figure~\ref{fig:attenuation_curve}). Dust has low abundance within 100 pc \citep{Lehner2003} and does not contribute significantly to the EUV opacity of the LISM. Despite significant ISM attenuation toward even the nearest stars, hundreds of sources have been detected in the short wavelength portion of the EUV by ROSAT/WFC \citep{Pye1995}, EUVE \citep{Bowyer1996}, and even Chandra \citep{Wood2018,Drake2020}.

\begin{figure}
    \centering
    \includegraphics[width=0.5\textwidth]{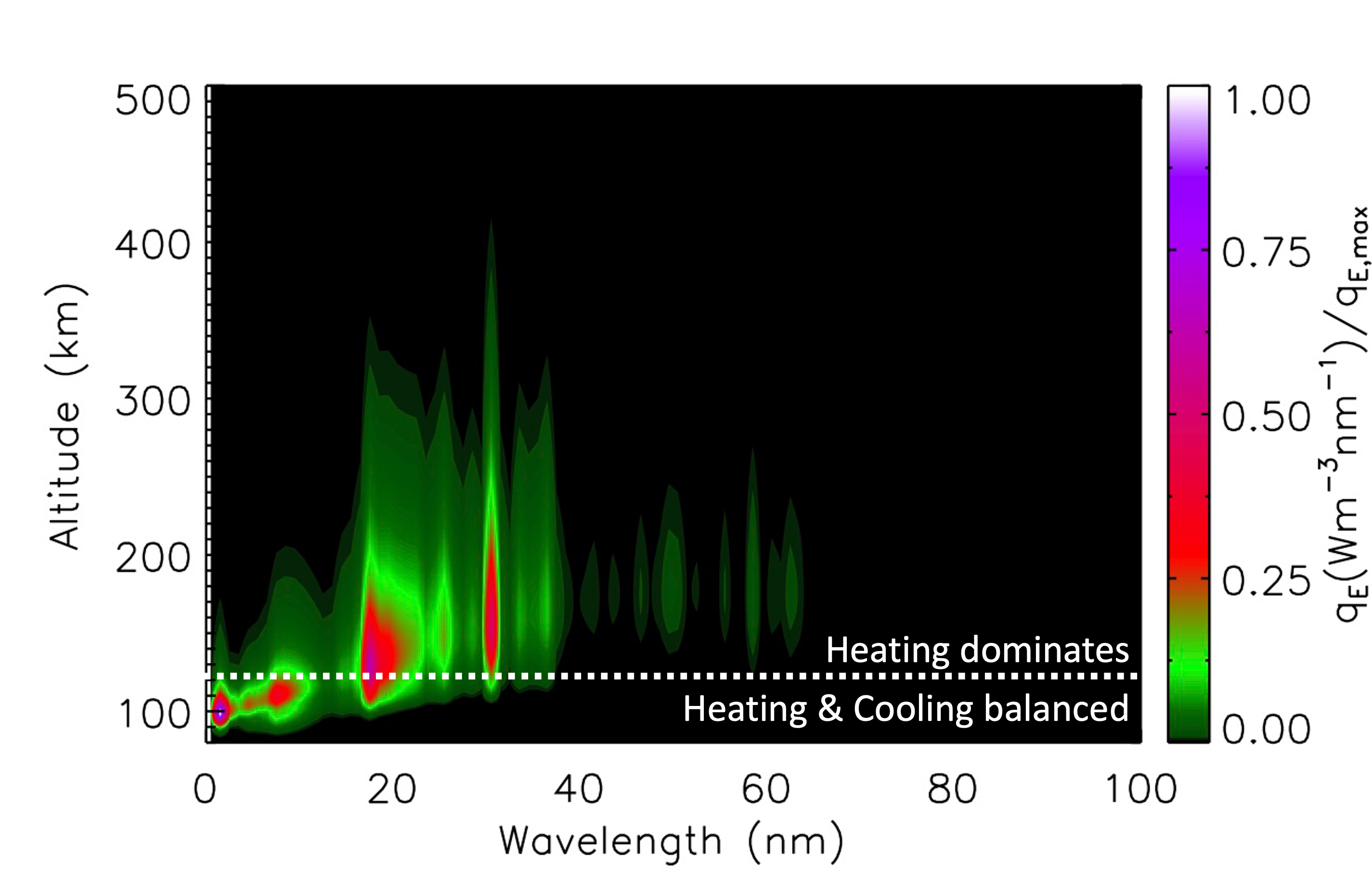}
    \caption{
    The contour plot shows, for the Earth's atmosphere, the specific energy input ($q_{\rm E}$) of  primary photoelectrons normalized by the peak of this energy input ($q_{\rm E, max}$) as a function of wavelength and altitude. This would represent the photoelectron heating rate for 100\% heating efficiency. In reality, the heating efficiency depends on atmospheric properties. An NRLMSIS atmosphere at 60$^{\circ}$ solar zenith angle and noon local time and a TIMED/SEE solar spectrum at moderate activity were used. The horizontal dotted line at $\sim$120 km demarcates the altitude above which heating dominates and escape occurs. This figure illustrates the relative importance of different wavelength regions to the EUV energy input.}
    \label{fig:photoelectrons}
\end{figure}

Since the EUVE mission concluded over 20 years ago, interest in the EUV spectral emission from cool stars has grown apace with the exoplanets field. EUV photons emitted from a host star are absorbed high in the atmosphere of orbiting planets, photoionizing the gas and driving atmospheric escape. Atmospheric escape is an important physical process affecting the evolution of all types of planets, especially rocky and sub-Neptune worlds \citep{Lopez2012,Gronoff2020}. Figure~\ref{fig:photoelectrons} shows the dependence of the photoelectron energy production rate (a proxy for heating) on the wavelength of incident light and altitude in the Earth's atmosphere. A {Naval Research Laboratory Mass Spectrometer Incoherent Scatter radar (NRLMSIS)} model of Earth's atmosphere \citep{Emmert2020} and a moderately-active solar spectrum from TIMED/SEE\footnote{\url{https://lasp.colorado.edu/data/timed\_see}} \citep{Woods2000} were assumed. Despite the greater number stellar photons from 400-911 \AA, the 100-400 \AA\ photons produce more energetic photoelectrons and dominate the energy input at high altitudes where heating dominates over radiative cooling processes on the Earth ($>$120 km).

Recovering the intrinsic emission of sources detected in the EUV requires correcting for the ISM attenuation, which is challenging when the \HI, \HeI, and \HeII\ column densities are not known for a given line of sight. \HI\ column densities are derived from \HI\ and \DI\ \Lya\ medium- to high-resolution absorption spectroscopy of (mostly) late type stars (see \citealt{Wood2005} for an overview) or EUV spectra of hot stars such as white dwarfs (see \citealt{Dupuis1995}). \HeI\ and \HeII\ column densities can only be derived from EUV spectra of hot stars \citep{Dupuis1995, Wolff1999}. Note however that linking \DI, \HeI, and \HeII\ column densities with \HI\ column density often requires assumptions about the relative abundance of these species when direct measurements are not possible. The D/H ratio has been tightly constrained for the LISM \citep{Wood2004,Linsky2006}, while empirical constraints on the He/H ratio from EUV spectra of hot white dwarfs are loose \citep{Dupuis1995}. In situ measurements of the LISM immediately surrounding the Sun with the IBEX mission \citep{Bzowski2019} and photoionization models \citep{Slavin2008} provide constraints in good agreement with the EUV-derived He/H and He ionization fraction estimates.

No all-sky map of the \HI\ distribution in the LISM is currently available. After over 30 years of HST observations, there are almost 200 point-like samplings of the LISM for \HI\, albeit with large gaps of many tens of degrees between some sightlines. Resonance transitions of ions such as \MgII\ and \FeII\ are also sensitive to LISM gas and easier to observe than \Lya\ or the EUV and therefore could offer improved spatial coverage, but the link between these metals and \HI\ abundances from cloud to cloud is not well established. Uncertainties on \NHI\ values derived from $N$(\MgII) and $N$(\FeII) are of the order 0.30 dex \citep{Linsky2019}. \HI\ 21 cm emission is generally not sensitive to the LISM because it is confused with the brighter background, with the exception of some nearby, small cold clouds (e.g., \citealt{Peek2011}) that are thought to span only a few square degrees of the sky in total. Also, Lyman series spectra with FUSE have limited use for typical LISM column densities (see \citealt{French2021}). 

A 3D model of 15 discrete LISM clouds has been constructed to reliably compute the kinematics of these clouds for a given sightline\footnote{\url{http://lism.wesleyan.edu/LISMdynamics.html}} \citep{Redfield2008,Linsky2019}, and a similar 3D model for one of the 15 clouds' column densities, the Local Interstellar Cloud (LIC) is currently available\footnote{\url{http://lism.wesleyan.edu/ColoradoLIC.html}} \citep{Redfield2000}. 
Such 3D maps made with abundant \NaI\ measurements \citep{Lallement2014, Vergely2010, Vergely2001} are generally only sensitive to colder, more distant interstellar environments \citep{Welsh2010}, and are therefore excellent maps of the greater Galactic arm environment out to 1 kpc, but are not very sensitive to the ISM within 10 pc of the Sun. {The 3D model of the LISM constructed by \cite{Gry2014} treats the LISM as a single cloud and recovers the observed kinematics of the LISM. However, column density predictions are not available from this model.}

In this work, we combine \NHI\ measurements towards \ntargets\ stars inside \limitingdistance\ pc (Section~\ref{sec:measurements}) to create all-sky \HI\ column density maps of the LISM for a range of distance shells (Section~\ref{sec:skymap}), and analyze how uncertainties in \NHI\ propagate to uncertainty in an object's assumed intrinsic EUV flux after ISM correction (Section~\ref{sec:uncert}). We conclude in Section~\ref{sec:summary}.

\begin{figure*}
    \centering
    \includegraphics[width=\textwidth]{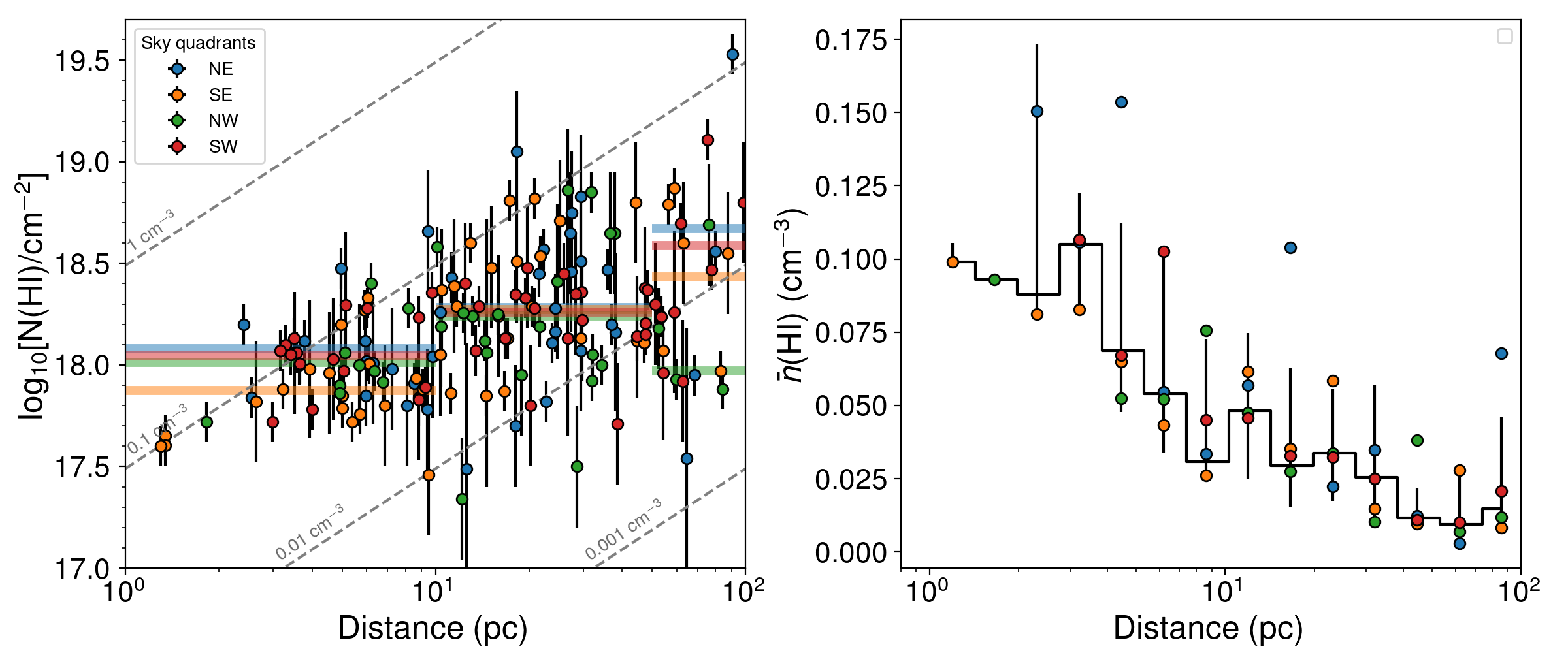}
    \caption{Left: The \HI\ column densities for each target in our sample as a function of distance. The targets are color-coded by sky quadrant {(see legend)}, where eastern targets have $l>180^{\circ}$ and northern targets have $b>0^{\circ}$. The gray dashed contours show constant, average line-of-sight \HI\ volume densities, \nhiavg, from upper left to lower right: 1.0, 0.1, 0.01, 0.001 cm$^{-3}$. Horizontal shaded lines show the average \NHI\ values for different distance shells {(1-10 pc, 10-50 pc, 50-100 pc)}, broken up by sky quadrant. Right: Median line-of-sight average volume densities are shown as a function of distance in logarithmic distance bins. Error bars show the inner 68\% spread of values centered around the median. Colored points show the median values for stars in the different sky quadrants.}
    \label{fig:NHI_distance}
\end{figure*}

\section{\HI\ column density measurements} \label{sec:measurements}

We assembled \NHI\ measurements of stars within \limitingdistance\ pc based on medium-to-high resolution \Lya\ observations ($R\sim$10,000-110,000) with the Hubble Space Telescope. {Most measurements come from the literature, and a dozen are drawn from works in preparation.} Note that when individual velocity component column densities were provided in the literature, we combined them to obtain a total \NHI\ value for that sightline. The \limitingdistance\ pc distance boundary was chosen because it is the rough extent of the Local Bubble. We focus on the GHRS and STIS instruments and exclude COS data because of airglow contamination, although see \cite{Bourrier2018_55Cnc} and \cite{CruzAguirre2023} for methods of removing this contamination from COS \Lya\ spectra. We elected not to use any \NHI\ measurements from EUV observations of hot white dwarfs. The vast majority of those measurements available in the literature are for stars beyond the \limitingdistance\ pc distance limit of this study and the methodology used for obtaining \NHI\ is different from the \Lya\ technique. 

The \HI\ \Lya\ line at 1215.67 \AA\ is the brightest emission line in the UV spectra of main sequence stars and serves as a strong backlight for foreground cold \HI\ and \DI\ gas in the intervening ISM. See \cite{Wood2005} and \cite{Youngblood2022} for comprehensive discussions on {how} the \HI\ column densities used in this paper were determined. In short, models of the \HI\ emission line representing the intrinsic stellar emission and \DI\ and \HI\ absorption lines representing the ISM attenuation are forward modeled to match the observed \Lya\ line profiles. The \HI\ absorption is heavily saturated even toward the nearest stars, and so the unsaturated \DI\ absorption line at 1215.34 \AA\ provides a significant constraint on the \HI\ column density. In many of the cases in Table~\ref{Tab:targs}, the mean D/H ratio for the LISM has been assumed, N(\DI)/\NHI\ = 1.5$\times$10$^{-5}$ \citep{Wood2004}, which is a good assumption given the small variation of the D/H ratio within the Local Bubble (1.56$\pm$0.04 $\times$10$^{-5}$; \citealt{Linsky2006}). 

In the {lower resolution spectra (R=10,000-20,000)}, the \DI\ and \HI\ lines are blended and cannot always be reliably disentangled. \cite{Wilson2022} demonstrate that the statistical uncertainties in \NHI\ {values derived from R$\sim$10,000 spectra} are underestimated, {possibly due to degeneracies between fitted parameters poorly accounted for by the fitting technique}. We have elected to include \NHI\ values from medium-resolution spectra despite this issue to greatly improve sky coverage. Many stars are faint at \Lya\ and inaccessible via high-resolution spectroscopy even with HST. Although the uncertainties from the medium-resolution spectra are larger, they can still demonstrate the rough \NHI\ parameter space such as strongly ruling out \NHI$>$10$^{19}$ cm$^{-2}$. 

The resulting number of targets used in our study is \ntargets.  Table~\ref{Tab:targs} lists target name, distance, coordinates, \NHI\ value, its uncertainty, instrument and grating, and the source of the \NHI\ value. {Column densities sourced from works in preparation are shown in Appendix~\ref{sec:appendix}.} We also list in Table~\ref{Tab:targs} the average \HI\ volume densities, discussed in the next subsection.

{Using such a broad sample from over 50 individual papers published by various groups over the span of three decades inevitably leads to inconsistencies in methods, assumptions, and the derivation of uncertainties. Reported statistical uncertainties usually do not capture larger systematic uncertainties due to instrument or data quality issues, degeneracies between fitted parameters (e.g., \citealt{Wilson2022}), assumptions made, or missing physics in the model. For example, most of the sample's reconstructions do not consistently treat the \Lya\ line's self-reversal \citep{Youngblood2022,Sandoval2023}, the LISM gas properties such as the D/H ratio (discussed above) or the temperature (e.g., the presence of hot gas; \citealt{Gry2017}), or the presence of astrospheric or heliospheric absorption \citep{Wood2005}. All of these assumptions can affect the measured \NHI\ in ways most likely not captured by the statistical uncertainties, some of which are smaller than 0.01 dex. It is beyond the scope of this paper to self-consistently measure \NHI\ values and determine appropriate uncertainties for the complete sample. 

To avoid giving individual sight lines with extremely small uncertainties undue influence on the derivation of the \NHI\ maps (Section~\ref{sec:skymap}), we have imposed a minimum uncertainty of 0.3 dex for all \NHI\ values derived from R=10,000-20,000 spectra (instrument modes: STIS/G140M, GHRS/G140M, GHRS/G160M) based on the analysis of \cite{Wilson2022}. We imposed a minimum uncertainty of 0.1 dex for the values derived from R$>$40,000 spectra (instrument modes: GHRS/ECH-A, STIS/E140M, STIS/E140H). For values without reported uncertainties, we assume the minimum uncertainty appropriate for the instrument mode. }

\subsection{LISM clouds probed by the average volume density} \label{subsec:volume_density}

Figure~\ref{fig:NHI_distance} shows as a function of distance the \NHI\ measurements and the inferred average \HI\ volume density, i.e., \nhiavg\ = \NHI/$d$ where $d$ is the star's distance in cm (see \citealt{Wood2005}). The average volume density \nhiavg\ assumes that the \HI\ completely fills the volume between the observer and the star and that the density is constant throughout this space. The true or intrinsic volume density \nhitrue\ of the material is unknown because the filling factor along the line of sight is generally unknown (except for the nearest sightlines within $\sim$4 pc, where it appears to be unity; \citealt{Linsky2022}) and the medium is inhomogeneous along a sightline \citep{Redfield2008,Gry2014}. Almost all \ntargets\ targets have values 17.5 $\leq$ \logNHI\ $\leq$ 19.0, roughly corresponding to \nhiavg$\sim$0.01-0.1 cm$^{-3}$. There appears to be a step-like increase in \NHI\ between approximately 1-10 pc, 10-50 pc, and 50-100 pc, with \nhiavg\ declining with increasing distance. This can be explained by low volume filling fractions for LISM \HI\ gas, with much of the \HI\ residing within $\sim$10 pc. The boundary of the Local Bubble is near $\sim$100 pc, possibly explaining the increase in \NHI\ near 100 pc. Various works have demonstrated that most of the LISM's detected \HI\ gas is nearby \citep{Redfield2004,Malamut2014} and has an intrinsic volume density \nhitrue $\sim$0.1 cm$^{-3}$ \citep{Linsky2022_AJ}. Beyond $\sim$15 pc and until the boundary of the Local Bubble, most interstellar hydrogen is ionized. 

\cite{Swaczyna2022} analyzed the mixing of {the LISM clouds from the \cite{Redfield2008} multi-cloud model} and provided tentative evidence that the Sun resides near the interaction region of two colliding clouds, the LIC and G clouds\footnote{{Under the \cite{Gry2014} LISM model, the LIC and the G clouds are two parts of the same cloud with a velocity gradient.}}. Compressed material in the interaction region should be detectable as enhanced \HI\ density, approximately in the shape of a ring encircling the sky position of the axis connecting the LIC and G cloud centers ($l,b$)=(334$^{\circ}$, 6$^{\circ}$). \cite{Swaczyna2022} identify AD Leo and LQ Hya (HD 82558) as two stars with volume density \nhiavg\ $>$ 0.1 cm$^{-3}$ near the expected location of the ring-like structure. The exact morphologies and extents of the LIC and G clouds are not known, so departures from a smooth or symmetric ring would be expected and the precise location of the LIC-G axis is not well constrained. 

\begin{figure*}
    \centering
    \includegraphics[width=\textwidth]{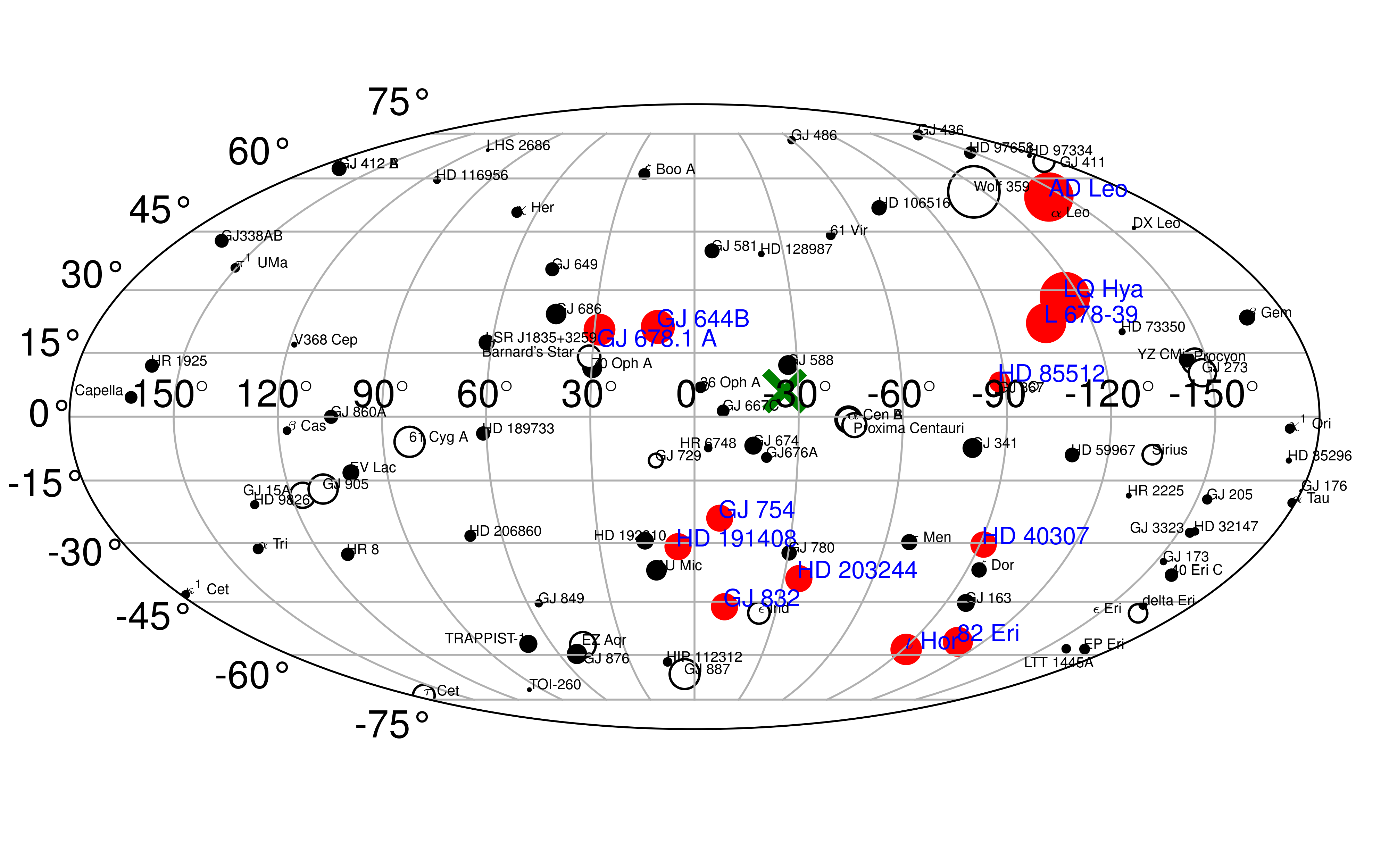}
    \caption{Map in Galactic coordinates of sight lines {within} 25 pc with \NHI\ values known to precision better than 0.3 dex. Stars within 4 pc are shown as open black circles, and stars between 4-25 pc are shown as filled circles. Stars beyond 4 pc with (\nhiavg\ + $\sigma_{\bar{n}(HI)}$) $>$ 0.1 cm$^{-3}$ are shown as filled red circles, otherwise filled black circles. The circle sizes are scaled linearly with the \nhiavg\ value of each sight line. The approximate position of the LIC and G cloud interaction axis is shown as a green X. }
    \label{fig:ring}
\end{figure*}

We use our dataset to explore if there are additional sight lines that support or refute the presence of a ring of enhanced density (Figure~\ref{fig:ring}). We used stars between 4-25 pc with {reported} \logNHI\ precisions better than 0.3 dex {(i.e., before the uncertainty adjustments described previously)}. The inner limit of 4 pc was selected because enhanced density due to compression should only be evident at distances greater than the thickness of the compression. The outer limit of 25 pc was selected because distant sight lines, even if they traverse the compression ring, will have low filling factors and the average volume density \nhiavg\ will be low. 
 We identified 15 stars with (\nhiavg\ + $\sigma_{\bar{n}}$) $>$ 0.1 cm$^{-3}$ that trace part of a ring-like structure around the LIC-G axis (red points in Figure~\ref{fig:ring}). \cite{Linsky2025} have similarly identified stars that may be part of the ring, and there is substantial overlap between the identified stars between these two works. We selected 0.1 cm$^{-3}$ for the threshold as it is the average volume density of the local clouds \citep{Linsky2022_AJ}.  There are several stars that challenge the ring hypothesis, including $\zeta$ Dor, GJ 163, and $\alpha$ Leo. The factor of two discrepancy between the $\zeta$ Dor and HD 40307 {average} density values is particularly surprising, since both stars are only 7$^{\circ}$ apart and are at about the same distance (12-13 pc). There are no sight lines near the north Galactic pole that exceed 0.1 cm$^{-3}$ (notably $\xi$ Boo A and GJ 486 have low volume densities), which could be because the solar system is too close to that region of the compression ring or the available sight lines simply miss the ring in that area. More sight lines are needed to {confirm} the compression hypothesis. However, further evidence for this {average} density enhancement being a LIC/G cloud compression region is that the (partial) ring seen in Figure~\ref{fig:ring} matches the G cloud boundary shown in the sky maps of \cite{Redfield2008}.

 It is worth pointing out that the highest mean volume densities, with \nhiavg $\sim$0.2 cm$^{-3}$, are observed in the same direction within the ring, roughly in the direction of the constellation of Leo.  In particular, four of the five sight lines with \nhiavg$>$0.15 cm$^{-3}$ are Wolf 359, AD Leo, LQ Hya, and L 678-39.  This general direction also includes the highest \NHI\ measurement within the Local Bubble, \logNHI\ = 19.53 for HR 3980.  Despite its distance of 90.7 pc, this star should still be well within the bounds of the Local Bubble according to the maps of \cite{Pelgrims2020} and \cite{O'Neill2024}.  This direction also includes the nearest cold H 21-cm emitting cloud, the Local Leo Cold Cloud (LLCC), which may be $\sim$20 pc away \citep{Peek2011,Meyer2012}.  \cite{Gry2017} note the similarity in motion between the LIC and the LLCC and suggest that they are related. Perhaps the LIC/G cloud interaction that we see locally is part of a larger scale flow interaction within the Local Bubble, which in the Leo direction results in the creation of the LLCC.

Compared to the high density sight lines that surround it, $\alpha$ Leo surprisingly has only a modest column density \logNHI=18.28, corresponding to an average volume density of only 0.025 cm$^{-3}$ (see Figure~\ref{fig:ring}).  {Through ionization modeling and electron density measurements, \cite{Gry2017} derived the intrinsic volume density of the LISM cloud toward this sight line to be \nhitrue=0.20$^{+0.08}_{-0.10}$ cm$^{-3}$, bringing this sight line into good agreement with the \nhiavg\ values of nearby stars such as AD Leo, and implying that the warm diffuse gas only extends $\sim$5 pc in this direction.}  We  note that this low column interpretation of the data requires the addition of an extra high temperature \HI\ absorption component to fit the \Lya\ data.  Without that component, the suggested column density could be as high as \logNHI=18.82, corresponding to \nhiavg=0.09 cm$^{-3}$. \cite{Gry2017} favored the low column model because the high column model would imply an unusually high depletion of metals towards $\alpha$ Leo. The Leo direction is worthy of more study in the future, at both near and far distances within the Local Bubble.

\section{An Interpolated Sky Map of \HI\ Column Density} \label{sec:skymap}

With the \texttt{tinygp} Python package version 0.2.0rc1 \citep{tinygp}, we used a Gaussian Process to spatially interpolate between the point-like \logNHI\ measurements to construct all-sky column density maps for various distance shells ($<$10 pc, 10-20 pc, 20-30 pc, 30-50 pc, 50-70 pc, 70-100 pc). We selected an exponential kernel described by three hyperparameters (scale distance, amplitude, and mean), and sampled the natural logarithm of these hyperparameters with the No U-Turn Sampler (NUTS) Markov Chain Monte Carlo (MCMC) algorithm via the Python package \texttt{NumPyro} \citep{phan2019composable,bingham2019pyro}. We selected prior probabilities on the hyperparameters to be relatively uninformative but centered around roughly expected values. We assumed a Cauchy prior of mean zero and scale 10 for the amplitude parameter (units: dex), a Gaussian prior of mean -1 and standard deviation 0.5 for the scale parameter (units: radians), and a Gaussian prior of mean 18.0 and standard deviation 2.0 for the average parameter (units: dex). Each measurement was weighted by the uncertainties reported in Table~\ref{Tab:targs}, and we used measurements inside/outside the distance shell as lower/upper limits by introducing a sigmoid function into the likelihood function. For example, for the 10-20 pc distance shell, we used all the \NHI\ values of targets $<$10 pc minus their uncertainties ($N$ - $\sigma_N$) as lower limits and the \NHI\ values plus their uncertainties ($N$ + $\sigma_N$) of targets $>$ 20 pc as upper limits. Convergence was assessed via the Gelman-Rubin statistic, which should be very close to unity. This kernel was constructed to mimic an exponential kernel used to spatially interpolate sparse temperature data in global maps (i.e., ordinary kriging; \citealt{pykrige}). 

\begin{figure*}
    \centering
    \includegraphics[width=\textwidth]{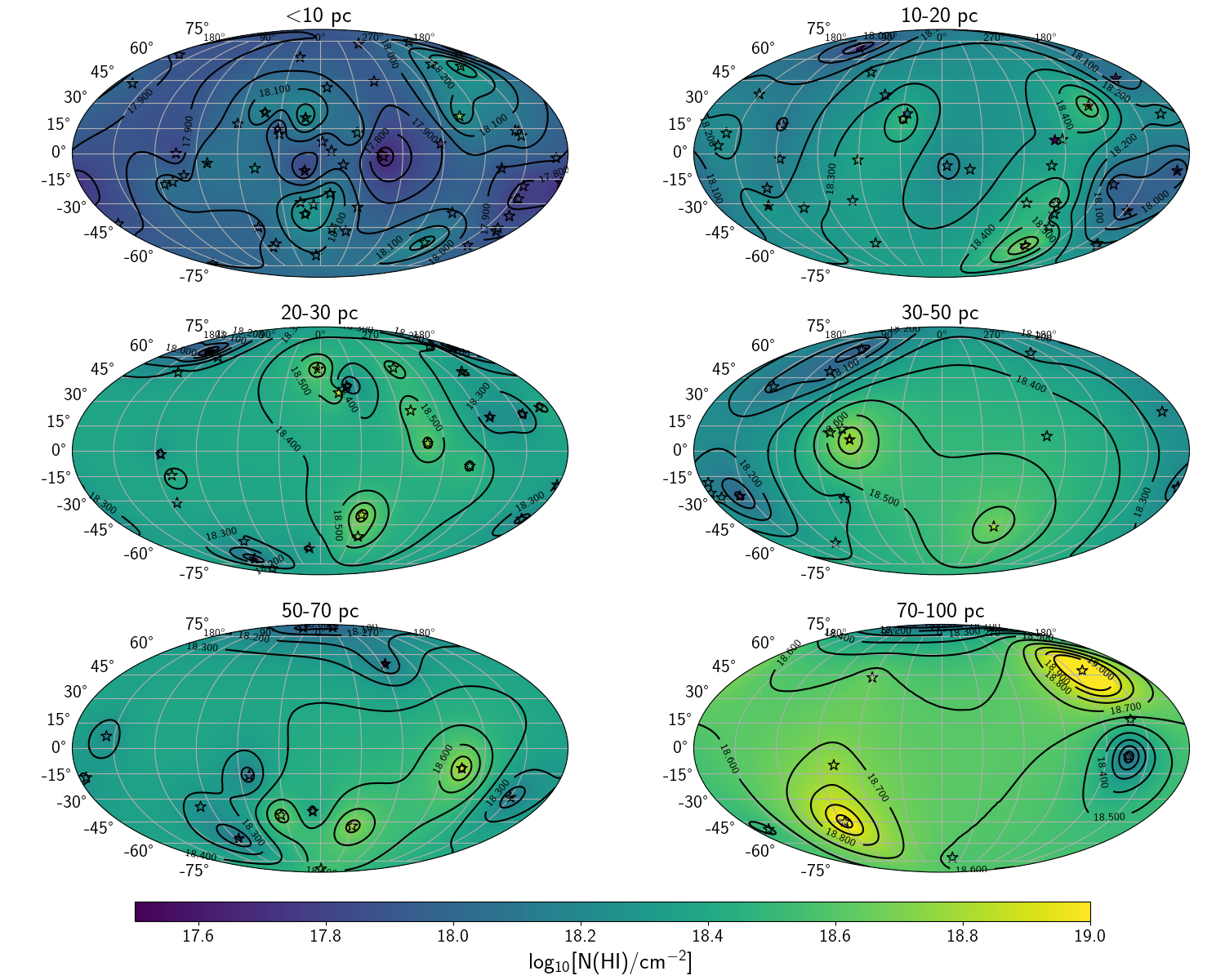}
    \caption{\HI\ column density maps are shown in Galactic coordinates for six different distance shells from Earth, from $<$10pc, up to 100 pc. The color shading shows the interpolated log$_{10}$ \NHI\ values based on the stars in that distance shell (marked as black-outlined stars with filled colors matching their measured \NHI\ values). The black contours show levels between \NHI=17.5-19.0 in 0.1 dex steps; they are printed with the value that they encircle. The star symbols are the observed stars in Table~\ref{Tab:targs}; the symbol colors represent the measured \NHI\ value.} 
    \label{fig:NHI_map}
\end{figure*}

Figure~\ref{fig:NHI_map} shows the derived column density maps for six different distance shells. The column density maps are publicly available in a public repository called \texttt{LISM\_NHI}\footnote{\url{https://github.com/allisony/LISM\_NHI}; \dataset[doi: 10.5281/zenodo.17087617]{\doi{10.5281/zenodo.17087617}}}. The code used to derive the maps is also publicly available\footnote{\url{https://github.com/allisony/LISM\_map}; \dataset[doi: 10.5281/zenodo.17087645]{\doi{10.5281/zenodo.17087645}}}.

The area-weighted median column densities across the entire sky are given in Table~\ref{Tab:averages}. We observe a {$\sim$0.4} dex increase in \logNHI\ from $<$10 pc to 10-20 pc, and a {$\sim$0.15} dex increase from 10-20 pc out to {50-70} pc. \logNHI\ further increases by {$\sim$0.15} dex in the outermost distance shell (70-100 pc). {Some of the area-weighted column densities differ from the average of the \NHI\ values input into the fit. For example, considering the target stars that fall within each of the six distance shells, we find mean values \logNHI\ = 17.99, 18.22, 18.35, 18.27, 18.24, and 18.62, in order of increasing distance. These mean values for the 20-30 pc, 30-50 pc, and 50-70 pc distance shells are 0.05-0.2 dex lower than the area-weighted averages reported in Table~\ref{Tab:averages}. This occurs because of a combination of two effects: the mean values of a distance shell do not account for lower/upper limits from the inner/outer distance shells, and the lower column sight lines in these distance shells are relatively spatially confined compared to the higher column sight lines. }

We recommend use of the standard deviation of the residuals (observed {N(HI)} - predicted {N(HI)}) as the uncertainty in {the \NHI\ estimates to avoid assigning them an unwarranted level of significance}. These uncertainties range from \lowerresidual-\upperresidual\ dex and are roughly inversely correlated with the number of stars in a shell (Table~\ref{Tab:averages}). Although using a uniform uncertainty for an entire map is an oversimplification because the statistical uncertainties are smaller nearer to the input measurements, the residuals more accurately represent the true uncertainty in these interpolated \logNHI\ values. 

{There is significant heterogeneity in the maps, with some ``holes" in the LISM apparent even at large distances, and two anomalously high \NHI\ sight lines in the 70-100 pc distance shell (GD 246 and HR 3980) that may be traversing the boundary of the Local Bubble. The hydrogen hole observed by \cite{Linsky2019} spans tens of degrees around 225$^{\circ} < l < 290^{\circ}$ and -60$^{\circ} < b < +10^{\circ}$ and can be seen in the southwest quadrant of the $<$10 pc map as a region of low column density. The hydrogen hole is driven by photoionization from hot stars, specifically $\epsilon$ CMa (l=239.8$^{\circ}$, b=-11.3$^{\circ}$, d=124.2 pc), $\beta$ CMa (l=226.1$^{\circ}$, b=-14.3$^{\circ}$, d=151.1 pc), and Sirius (l=227.2$^{\circ}$, b=-8.9$^{\circ}$, d=2.6 pc) \citep{Linsky2019}. The maps do not include $\epsilon$ CMa or $\beta$ CMa because they are beyond 100 pc. From EUV spectra of $\epsilon$ CMa, \cite{Shull2025} derived \logNHI=17.8$\pm$0.1 dex. The 70-100 pc map predicts \logNHI = 18.38$\pm$0.48 toward $\epsilon$ CMa, which is substantially higher than the value from \cite{Shull2025}. Therefore, our maps are likely not capturing the true extent of the hydrogen hole. 

The unusual $\epsilon$ CMa sightline provides a useful demonstration of the maps and their limitations. While the outermost distance shell predicted a \NHI\ value substantially larger than the EUV-calculated value, they were in agreement within uncertainties. By comparison, the 50-70 pc distance shell predicts a higher value (18.56$\pm$0.42) than the 70-100 pc shell and is not consistent within uncertainties with the EUV-derived value. This discrepancy between an inner/outer shell is allowed by the inclusion of measurement uncertainties in the definition of the lower/upper limits that were imposed on the fit by the inner/outer distance shells. Additionally, some outliers are expected in the fitting process. Another factor at play, especially in the outer distance shells, is the number of stars informing the fit in each area of the sky. Individual stars strongly dominate the contours of the maps within a few degrees. In the 50-70 pc shell, the nearest neighbor to the $\epsilon$ CMa sightline is $\sigma$ Pup (l=255.7$^{\circ}$, b=-11.9$^{\circ}$, d=58.7 pc) which has a relatively high \NHI\ value (18.87; \citealt{Wood2024}) that is increasing that shell's \NHI\ estimate toward $\epsilon$ CMa.  Far from individual stars, the map values tend toward the area-weighted average for that distance shell.

Additional \Lya\ measurements in sky regions with few measurements at a wide range of distances within $\sim$100 pc would improve our understanding of the distribution of \HI\ within the Local Bubble, including the hydrogen hole. Self-consistent reconstructions of the set of available \Lya\ spectra toward nearby stars could also improve the maps.} 

\begin{deluxetable}{cccc}
    \tablecolumns{4}
\tablewidth{0pt}
\tablecaption{Results of sky interpolation. \label{Tab:averages}}
\tablehead{\colhead{Distance} &
                  \colhead{Median} &
                  \colhead{Std. Dev. of} &
                  \colhead{Number of} \\
                  \colhead{shell} &
                  \colhead{\logNHI} &
                  \colhead{Residuals (dex)} &
                  \colhead{Stars}
                  }
\startdata
$<$10 pc & {17.87$\pm0.07$} & {0.20} & 61 \\
10-20 pc & {18.25$^{+0.07}_{-0.04}$} & {0.36} & {34} \\
20-30 pc & {18.39$^{+0.06}_{-0.07}$} & {0.35} & 28 \\
30-50 pc & {18.47$^{+0.11}_{-0.09}$} & {0.36} & {18} \\
50-70 pc & {18.45$^{+0.11}_{-0.09}$} & {0.42} & 14 \\
70-100 pc & {18.62$\pm$0.22} & {0.48} & 9 \\
\enddata
\tablecomments{The first column indicates the distance shells for which we performed a sky interpolation. The second column shows the average hyperparameter's median and 68\% confidence interval. The third column shows the residuals which were calculated as {the standard deviation of the} observed \logNHI\ minus the predicted value from the interpolated sky map for that distance shell. The fourth column shows the number of stars that fell in the distance shell. Upper/lower limits were included on all fits for stars exterior/interior to the distance shell.}
\end{deluxetable}

\section{The impact of \NHI\ uncertainty on intrinsic UV flux measurements} \label{sec:uncert}

\subsection{EUV} \label{subsec:EUV_uncert}

EUV photons are absorbed by LISM H and He gas, with substantial attenuation even for the lowest column densities observed. This is due to the large bound-free cross sections of \HI, \HeI, and \HeII\ through the EUV and the abundance of these gases in the LISM. The wavelength-dependent opacity $\tau_{\lambda}$ for each species is given as the product of the bound-free cross section $\sigma_{bf}$ (cm$^{2}$) and the species' column density $N$ (cm$^{-2}$). For hydrogenic atoms and ions like \HI\ and \HeII, the bound-free cross sections are given by:

\begin{equation}
    \sigma_{bf}(\lambda) = \frac{7.91\times 10^{-18} \rm{cm}^2}{Z^2} \cdot \Big(\frac{\lambda}{\lambda_0}\Big)^3 \cdot g_{bf}(\lambda)
\end{equation}

\noindent where $Z$ is the electric charge ($Z$=1 for \HI\ and $Z$=2 for \HeII), $\lambda$ is the wavelength in \AA, and $\lambda_0$ is the Doppler-shifted ionization edge for the species ($\lambda_0$=912 \AA, 228 \AA\ for \HI\ and \HeII, respectively (under the assumption of RV=0 \kms)\footnote{Unlike bound-bound cross-sections, bound-free cross-sections are slowly varying with wavelength. RV $<$100 \kms\ results in a $<$0.3 \AA\ Doppler shift, which is not a large impact. In this work, we assume RV=0 \kms\ for simplicity.}, and $g_{bf}$ is the wavelength-dependent Gaunt factor, a quantum mechanics correction term applied to bound-free transitions \citep{Spitzer1978}.  For \HeI, we use the bound-free cross sections as a function of wavelength from \cite{Samson1966}. The column densities $N$ for \HeI\ and \HeII\ were linked to \NHI\ as follows: N(\HeI)/\NHI\ = 0.08 and N(\HeII)/\NHI = 0.12 \citep{Dupuis1995}. Note that the \HeII\ fraction is actually an upper limit and is based on the $<$60\% ionization fraction of He provided by \cite{Dupuis1995}. The opacities for \HI, \HeI, and \HeII\ are summed to obtain the total opacity as a function of wavelength. The probability of a photon at wavelength $\lambda$ passing through the medium is $e^{-\tau_{total}(\lambda)}$, and we refer to this term as the ISM attenuation or transmission fraction. 

Figure~\ref{fig:attenuation_curve} shows example absorption curves for the bound-free transitions described above. Note that there are additional bound-bound transitions of \HI, \HeI, and \HeII\ that contribute to opacity throughout the UV. The locations of these transitions are shown as tick marks in the figure, and a detailed example of the transmission fraction for \Lya\ at 1215.67 \AA\ is shown in the right panel (see Section~\ref{subsec:LyA}). 

\begin{figure*}
    \centering
    \includegraphics[width=0.95\textwidth]{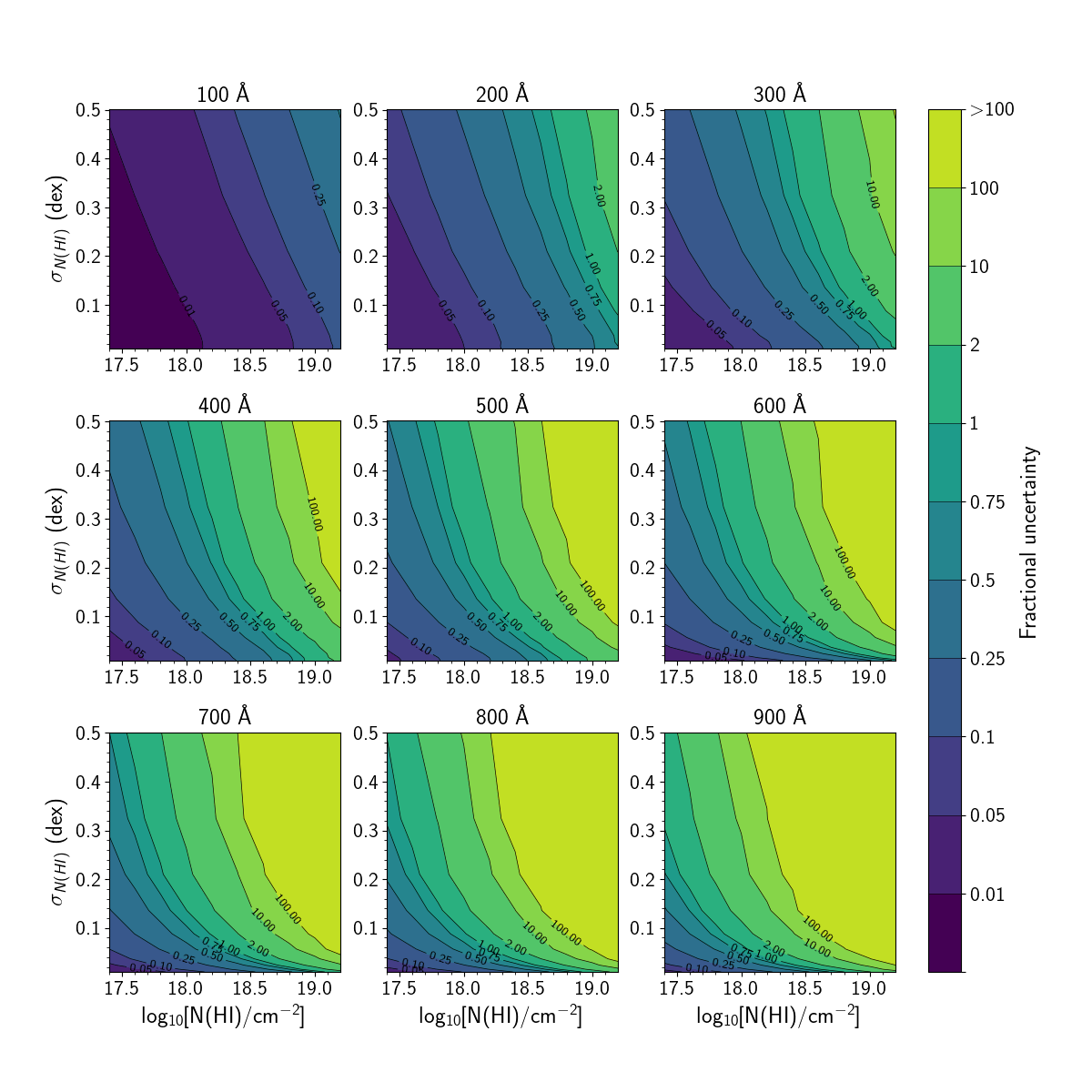}
    \caption{The fractional uncertainty of intrinsic EUV flux post-ISM attenuation correction is shown as a function of \logNHI\ value and uncertainty on that value at nine discrete wavelengths between 100-900 \AA. Bound-free (continuum) absorption from interstellar \HI, \HeI, and \HeII\ dominates at these specific wavelengths. We have included uncertainty on the He abundance (0.08$\pm$0.02) and the upper limit on the He ionization ($<$0.60) from \cite{Dupuis1995} in the simulations.} 
    \label{fig:fractional_uncertainties_EUV}
\end{figure*}

To understand how uncertainty in \HI\ column density translates to uncertainty in the assumed ISM attenuation curve in the EUV, we performed calculations of ISM attenuation over a grid of \logNHI\ values and uncertainties ($\sigma_{N(HI)}$) between 17.4 and 19.2 dex and 0.01-0.5 dex, respectively. These ranges fully encompass the observed \HI\ column densities inside 100 pc and the measurement errors typically obtained via \Lya\ reconstructions or the interpolated maps of Section~\ref{sec:skymap}. For the $i$th grid cell, we obtained 10,000 plausible values of \logNHI\ by randomly drawing from a normal distribution of mean \logNHI$_i$\ and standard deviation $\sigma_{N(HI),i}$. Simultaneously, we also made random draws for the He/H and \HeII/\HeI\ fractions. We assumed a normal distribution of mean 0.08 and standard deviation 0.02 for He/H and a uniform distribution between 0-0.6 for \HeII/\HeI\ to represent the upper limit nature of the constraint from \cite{Dupuis1995}. We used these randomly drawn \HI, \HeI, and \HeII\ column densities to compute 10,000 ISM attenuation curves. Note that we only considered bound-free (continuum) absorption by H and He as this dominates the LISM's opacity in the EUV. {We calculate the fractional uncertainty on the attenuation factor at every wavelength bin as the 68th interpercentile range (from the 16th to the 84th percentile) in the spread of attenuation factors normalized by the median attenuation factor (50th percentile). The fractional uncertainty is defined as $(e^{-\tau}_{84.1\%} - e^{-\tau}_{15.9\%}) / e^{-\tau}_{50\%}$, where $e^{-\tau}$ represents the attenuation factor in each wavelength bin and the subscripts denote the percentile value.} Figure~\ref{fig:fractional_uncertainties_EUV} shows these uncertainties in each grid cell at wavelengths 100-900 \AA\ in steps of 100 \AA. As expected, larger \NHI\ uncertainties translate to larger EUV flux uncertainties at longer wavelengths where the opacity is largest. The fractional uncertainties do exceed unity in some of the regions of parameter space with higher \NHI\ uncertainty and/or longer wavelengths, indicating that a corrected EUV flux in those regimes would be {uncertain} to within a factor of two or greater.

The interpolated \logNHI\ maps presented in Section~\ref{sec:skymap} have uncertainties in the \lowerresidual-\upperresidual\ dex range. Here we present a specific scenario when using the all-sky maps. For a sightline at 8 pc near {(l, b) = (110$^{\circ}$, -30$^{\circ}$)}, the interpolated $<$10 pc map returns \logNHI\ = {18.02$\pm$0.20} dex. The expected uncertainty on intrinsic EUV fluxes due to the ISM attenuation correction alone would be  $\sim$5\% at 100 \AA, $\sim$25\% at 300 \AA, $\sim$100\% (i.e., a factor of two) at 500 \AA, and $>$100\% between 650-912 \AA. As seen in Table~\ref{Tab:targs}, direct \Lya\ determinations of \NHI, especially from high-resolution spectra, have statistical uncertainties of $\sim$0.05 dex or less, which significantly improves the EUV intrinsic flux uncertainty. For a similar \logNHI\ = 18.0 case, 0.05 dex uncertainty corresponds to $\sim$1\% at 100 \AA, $\sim$10\% at 300 \AA, $\sim$25\% at 500 \AA, and maintains $<$100\% uncertainty up to 900 \AA. 

As Figure~\ref{fig:photoelectrons} demonstrates, EUV observations in the 100-400 \AA\ are the highest priority region for understanding atmospheric escape from exoplanets. Fortunately, uncertainties from the ISM correction are minimized in this wavelength region.

\begin{figure*}
    \centering
    \includegraphics[width=0.95\textwidth]{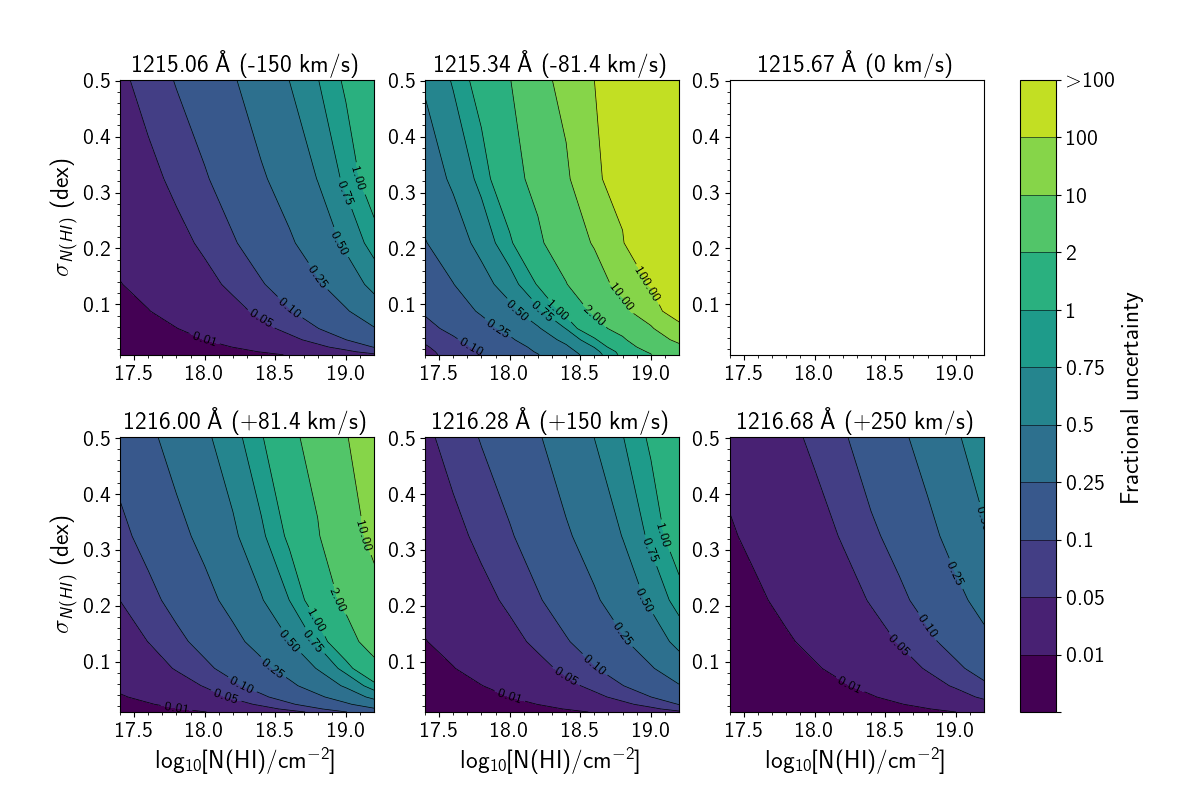}
    \caption{The fractional uncertainty of intrinsic \Lya\ flux post-ISM attenuation correction is shown as a function of \logNHI\ value and uncertainty on that value at six discrete wavelengths across the \HI\ \Lya\ line. Near the line center (1215.67 \AA\ or 0 \kms), the ISM transmission is 0\% regardless of the \NHI\ value, so the fractional uncertainty is undefined. We have included uncertainty in the \DI\ abundance (D/H = 1.56$\pm$0.04 $\times$10$^{-5}$) and Doppler broadening parameter for \HI\ ($b$ = 10.8$\pm$1.0 \kms) in the simulations.} 
    \label{fig:fractional_uncertainties_Lya}
\end{figure*}

\subsection{\Lya} \label{subsec:LyA}

We conduct a similar exercise for the \HI\ \Lya\ line at 1215.67 \AA\ to understand the impact of \NHI\ uncertainty on the apparent flux density of emission near this transition. The \Lya\ line is the brightest UV emission line of cool stars and is a useful tracer of stellar chromospheres and their photochemical impact on orbiting exoplanets (e.g., \citealt{Wood2005,Youngblood2022,Peacock2022}) as well as probing escaping hydrogen from exoplanet atmospheres (e.g., \citealt{dosSantos2020}). \Lya\ is also vital for studies of the interstellar medium and was the basis for all of the interstellar column density measurements used in this work (Section~\ref{sec:measurements}). The high opacity of interstellar \HI\ and \DI\ at 1215.67 and 1215.34 \AA, respectively, generates substantial, wavelength-dependent attenuation near 1216 \AA, with 100\% of photons absorbed in the \HI\ line core, even under the lowest column density conditions ever observed in the LISM. This substantial interstellar attenuation complicates observations of this important transition. 

We use the \texttt{lyapy} code\footnote{\url{https://github.com/allisony/lyapy}; \dataset[doi: 10.5281/zenodo.15035159]{\doi{10.5281/zenodo.15035159}}} \citep{lyapy} to model the cross-sections of \HI\ and \DI\ absorbers as Voigt profiles. We assumed mean rest wavelengths, oscillator strengths, and natural damping constants for these hyperfine doublets from \cite{Morton2003}. Figure~\ref{fig:attenuation_curve} shows example attenuation curves across the \Lya\ line for a the range of \HI\ column densities observed in this LISM; other parameters are fixed to typical values: radial velocity (RV=0 \kms), \HI\ Doppler broadening parameter ($b_{HI}$=10.8 \kms), and D/H ratio (D/H=1.56$\times$10$^{-5}$). The \DI\ Doppler broadening parameter is fixed relative to the \HI\ parameter: $b_{DI}$ = $b_{HI}$/$\sqrt{2}$. The \HI\ and \DI\ wavelength-dependent opacities are computed as the product of the bound-bound cross-sections and the species' column density and summed ($\tau_{total}$ = $\tau_{HI}$ + $\tau_{DI}$). The transmission fraction is $e^{-\tau_{total}}$.

We calculated the ISM attenuation over the same grid of \logNHI\ values and uncertainties ($\sigma_{N(HI)}$) as Section~\ref{subsec:EUV_uncert}.  For the $i$th grid cell, we obtained 10,000 plausible values of \logNHI\ by randomly drawing from a normal distribution of mean \logNHI$_i$\ and standard deviation $\sigma_{N(HI),i}$. Simultaneously, we also made random draws for the $b_{HI}$ and D/H parameters. We assumed a normal distribution with mean 10.8 \kms\ and standard deviation 1.0 \kms\ for $b_{HI}$ \citep{Redfield2004} and a normal distribution with mean 1.56$\times$10$^{-5}$ and standard deviation 0.04$\times$10$^{-5}$ for D/H \citep{Wood2004,Linsky2006}. We used these randomly drawn properties to compute 10,000 ISM attenuation curves. We determined the fractional uncertainty at every wavelength bin in the same way described in Section~\ref{subsec:EUV_uncert}. Figure~\ref{fig:fractional_uncertainties_Lya} shows these uncertainties in each grid cell at six discrete positions along the \Lya\ line, given in velocities as -150 \kms, -81.4 \kms\ (\DI\ absorption is centered here), 0 \kms\ (\HI\ absorption is centered here), +81.4 \kms, +150 \kms, and +250 \kms.

The fractional uncertainties are undefined at 0 \kms\ because the ISM transmission fraction is zero regardless of the \NHI\ value. The fractional uncertainties decline with increasing separation from the \HI\ line center, except for an increase in uncertainty at the \DI\ line center (-81.4 \kms). The interpolated sky map and the uncertainty estimates may be useful for planning \Lya\ observations, especially of transiting exoplanets where stellar emission $\approx$ -200 to -100 \kms\ from line center are commonly used as a backlight for probing escaping hydrogen (e.g., \citealt{Ehrenreich2015}). For example, for a sightline at 8 pc near {(l, b) = (110$^{\circ}$, -30$^{\circ}$)}, the interpolated $<$10 pc map returns \logNHI\ = {18.02$\pm$0.20} dex. The uncertainty in the apparent flux density at -150 \kms\ is between 5-10\%, while at -81.4 \kms, it is near 100\% (i.e., a factor of two uncertainty). Note that once \Lya\ observations are in hand, standard reconstruction techniques can be used to recover the intrinsic \Lya\ profile and integrated line flux, along with \HI\ and \DI\ interstellar absorption properties (e.g., \citealt{Wood2005,Youngblood2022}).

\section{Summary and Conclusions} \label{sec:summary}
We have presented new all-sky \HI\ column density maps created by spatially interpolating point-like \NHI\ measurements from \HI\ \Lya\ toward \ntargets\ individual sight lines within \limitingdistance\ pc. We have confirmed the finding from past studies that within \limitingdistance\ pc (the confines of the Local Bubble), the range of observed column densities is 17.5 $\leq$ \logNHI\ $\leq$ 19.0, implying average volume densities of \HI\ between 0.01-0.1 cm$^{-3}$ \citep{Wood2005}. These {average} volume densities decrease from $\sim$0.1 cm$^{-3}$ near the Sun to $\sim$0.01 cm$^{-3}$ at 100 pc, resulting in an approximately flat average \NHI\ distribution with distance between 10-100 pc. From our interpolated sky maps created for different distance shells, sightlines $<$10 pc have an average \logNHI\ $\sim$ {17.9} and further out, values increase to {18.3} (10-20 pc), 18.4 (20-70 pc), and 18.6 (70-100 pc). This indicates that the vast majority of the \HI\ gas in the Local Bubble is within 10-20 pc of the Sun, in agreement with previous studies (\citealt{Redfield2004,Redfield2008,Malamut2014,Linsky2022,Linsky2025}. There is good agreement between the distance shells $>$10 pc, with a slight increase in \NHI\ for the outermost distance shell. The outer boundary of the Local Bubble is not exactly at 100 pc in every direction \citep{Welsh2010,Zucker2022,Lallement2022}, so some stars' column densities in this outermost shell may include \HI\ outside the Local Bubble. We also examine evidence for a ring of compressed gas tracing the collision between the LIC and G clouds of the LISM \citep{Swaczyna2022}, confirming tentative evidence of such a structure in the expected sky location.

We estimate that the interpolated sky maps have uncertainties in the range of \lowerresidual-\upperresidual\ dex. These uncertainties are roughly inversely proportional to the number of stars residing in the distance shell.  For comparison, directly-measured \NHI\ values from high-resolution \Lya\ spectroscopy where the \DI\ absorption line is well-resolved tend to have statistical uncertainties smaller than 0.05 dex. For medium-resolution spectra where the \DI\ line is not well-resolved, statistical uncertainties are larger, typically 0.10-0.50 dex. {Systematic uncertainties, discussed briefly in Section~\ref{sec:measurements}, are harder to quantify and are most likely larger than the statistical uncertainties. The reader should avoid over-interpreting the significance of the map-derived \NHI\ estimates, especially for the outer distance shells with few sight lines, because the map contours can be strongly dominated by individual stars. However, it is common practice to estimate \NHI\ based on the values of stars that are nearby in projection, and the presented maps provide a more rigorous way to make those estimates.}

We calculated how uncertainties in \NHI\ propagate to uncertainty in intrinsic EUV and \Lya\ fluxes after correction for ISM attenuation. These results combined with the \NHI\ sky maps can be used for planning observations of targets inside \limitingdistance\ pc at \Lya\ or EUV wavelengths where ISM attenuation is strong. For example, when planning \Lya\ observations with HST, the sky maps can generate an \NHI\ estimate with corresponding uncertainty based on the coordinates and distance of the star. These values and their uncertainties can be used with Figure~\ref{fig:fractional_uncertainties_Lya} to explore the likely range of expected interstellar attenuation across the \Lya\ line, under the assumption that the interstellar absorption is centered at 0 \kms. The kinematic model of \cite{Redfield2008} can provide a radial velocity estimate for any sight line (see footnote 2). Similarly, for EUV observation planning, \NHI\ and uncertainty estimates for individual targets are provided by the sky maps, and Figure~\ref{fig:fractional_uncertainties_EUV} provides the likely range of expected interstellar attenuation throughout the EUV. The radial velocity of the absorbers is not important to include for the EUV given the relatively slow changes of transmission factor with wavelength across the very broad EUV bandpass compared to the rapid changes across the narrow \Lya\ wavelength region.

Obtaining new \Lya\ measurements of more sightlines would improve the \NHI\ estimates presented here. Most of the observations utilized here were not made for the purposes of studying the LISM; therefore the coverage of the sky is piecemeal and not a uniform or comprehensive sampling of sky location and distance. While simply filling gaps in the sky coverage of \NHI\ measurements would greatly improve the estimates of the morphological distribution of nearby \HI\ gas, a more powerful approach would be to combine the kinematic and morphological information available for these sight lines to generate a physically motivated model of LISM gas. \cite{Redfield2000} presented such a model for a single LISM cloud (the Local Interstellar Cloud), but there {could be} at least 14 other kinematically-distinct clouds \citep{Redfield2008,Linsky2019}, {although see \cite{Gry2014} for an alternate kinematic model consisting of a single LISM cloud that is non-uniform}. High-resolution UV spectroscopy distinguishes the absorption lines of individual {cloud components}, which each have distinct radial velocities but often spatially overlap in projection on the sky, allowing the construction of a 3D model for each cloud {component}. Such a model should also have greater predictive power than the maps presented here for sky regions with no prior \NHI\ constraints.

\begin{acknowledgments}
{We thank the anonymous referee for providing suggestions that significantly improved the paper.} A.Y. thanks Daniel Foreman-Mackey for assistance with the \texttt{tinygp} package and Kosuke Namekata for providing the \HI\ column density of EK Dra. SR and JL thank the NASA Outer Heliosphere Guest Investigators Program for support to Wesleyan University and the University of Colorado for grant 80NSSC20K0785. This project has received funding from the European Research Council (ERC) under the European Union's Horizon 2020 research and innovation programme (project {\sc Spice Dune}, grant agreement No 947634). This research has made use of the SIMBAD database, CDS, Strasbourg Astronomical Observatory, France. 
\end{acknowledgments}

\software{Astropy \citep{Robitaille2013}, IPython \citep{Perez2007}, Matplotlib \citep{Hunter2007}, NumPy and SciPy \citep{VanderWalt2011}, NumPyro \citep{phan2019composable,bingham2019pyro}, Pandas \citep{Pandas}, JAX \citep{Jax}, tinygp \citep{tinygp}, lmfit \citep{lmfit}, lyapy \citep{lyapy}, LISM\_NHI \citep{LISM_NHI}}

\appendix
\section{Targets and Reconstructions} \label{sec:appendix}

\bibliography{main_arxiv.bbl}{}
\bibliographystyle{aasjournalv7}

\startlongtable
\begin{deluxetable*}{cccccccccccc}
\tablecolumns{12}
\tablewidth{0pt}
\tablecaption{Target list and \NHI\ values \label{Tab:targs}}
\tablehead{\colhead{Star} &
                  \colhead{l} &
                  \colhead{b} &
                  \colhead{d} &
                  \colhead{log$_{10}$[} &
                  \colhead{$\sigma_{N(HI)}$} &
                  \colhead{$\sigma_{N(HI)}^{\rm adj.}$} &
                  \colhead{\nhiavg} &
                  \colhead{$\sigma_{\bar{n}(HI)}$} &
                  \colhead{$\sigma_{\bar{n}(HI)}^{\rm adj.}$} &
                  \colhead{Instrument/} &
                  \colhead{Ref.} \\
                  \colhead{Name} &
                  \colhead{(deg)} &
                  \colhead{(deg)} &
                  \colhead{(pc)} & 
                  \colhead{N(HI)/cm$^{-2}$]} &
                  \colhead{(dex)} &
                  \colhead{(dex)} &
                  \colhead{(cm$^{-3}$)} &
                  \colhead{(cm$^{-3}$)} &
                  \colhead{(cm$^{-3}$)} &
                  \colhead{Grating} &
                  \colhead{}
                  }
\startdata
Proxima Centauri & 313.94 & -1.93 & 1.3 & 17.6 & 0.0 & 0.1 & 0.1 & 0.0 & 0.02 & STIS/E140M & 1 \\
$\alpha$ Cen B & 315.73 & -0.68 & 1.35 & 17.6 & 0.007 & 0.1 & 0.1 & 0.002 & 0.02 & GHRS/ECH-A & 2 \\
$\alpha$ Cen A & 315.73 & -0.68 & 1.35 & 17.65 & 0.01 & 0.1 & 0.11 & 0.003 & 0.02 & GHRS/ECH-A & 2 \\
Barnard's Star & 31.01 & 14.06 & 1.83 & 17.72 & 0.03 & 0.1 & 0.09 & 0.006 & 0.02 & STIS/E140M & 3 \\
Wolf 359 & 244.05 & 56.12 & 2.41 & 18.2 & 0.0 & 0.1 & 0.21 & 0.0 & 0.05 & STIS/E140M & 4 \\
GJ 411 & 185.12 & 65.43 & 2.55 & 17.84 & 0.03 & 0.1 & 0.09 & 0.006 & 0.02 & STIS/E140M & 3 \\
Sirius & 227.23 & -8.89 & 2.64 & 17.82 & 0.09 & 0.3 & 0.08 & 0.02 & 0.06 & GHRS/G140M & 5 \\
GJ 729 & 11.31 & -10.28 & 2.98 & 17.72 & 0.03 & 0.1 & 0.06 & 0.004 & 0.01 & STIS/E140M & 6 \\
GJ 905 & 109.99 & -16.94 & 3.16 & 18.07 & 0.0 & 0.1 & 0.12 & 0.0 & 0.03 & STIS/E140M & 4 \\
$\epsilon$ Eri & 195.84 & -48.05 & 3.22 & 17.88 & 0.07 & 0.1 & 0.08 & 0.01 & 0.02 & GHRS/ECH-A & 7 \\
GJ 887 & 5.1 & -65.96 & 3.29 & 18.1 & 0.01 & 0.1 & 0.12 & 0.003 & 0.03 & STIS/E140M & 8 \\
EZ Aqr & 47.07 & -56.98 & 3.41 & 18.05 & 0.0 & 0.1 & 0.11 & 0.0 & 0.02 & STIS/E140M & 4 \\
61 Cyg A & 82.32 & -5.82 & 3.5 & 18.13 & 0.03 & 0.1 & 0.12 & 0.009 & 0.03 & GHRS/ECH-A & 9 \\
Procyon & 213.7 & 13.02 & 3.51 & 18.06 & 0.01 & 0.3 & 0.11 & 0.002 & 0.07 & GHRS/G160M & 10 \\
GJ 15A & 116.68 & -18.45 & 3.56 & 18.06 & 0.02 & 0.1 & 0.1 & 0.005 & 0.02 & STIS/E140M & 6 \\
$\epsilon$ Ind & 336.19 & -48.04 & 3.64 & 18.0 & 0.1 & 0.1 & 0.09 & 0.02 & 0.02 & GHRS/ECH-A & 11 \\
$\tau$ Cet & 173.1 & -73.44 & 3.65 & 18.01 & 0.002 & 0.1 & 0.09 & 0.0004 & 0.02 & STIS/E140M & 12 \\
GJ 273 & 212.34 & 10.37 & 3.79 & 18.12 & 0.02 & 0.1 & 0.11 & 0.005 & 0.03 & STIS/E140M & 8 \\
GJ 191 & 250.53 & -36.0 & 3.93 & 17.98 & 0.34 & 0.34 & 0.08 & 0.06 & 0.06 & STIS/E140M & 3 \\
GJ 860A & 104.69 & -0.0 & 4.01 & 17.78 & 0.02 & 0.1 & 0.05 & 0.002 & 0.01 & STIS/E140M & 8 \\
GJ 674 & 343.0 & -6.78 & 4.55 & 17.96 & 0.01 & 0.3 & 0.06 & 0.001 & 0.04 & STIS/G140M & 6 \\
GJ 876 & 52.0 & -59.63 & 4.67 & 18.03 & 0.04 & 0.3 & 0.07 & 0.007 & 0.05 & STIS/G140M & 13 \\
GJ 412 A & 168.5 & 63.07 & 4.9 & 17.9 & 0.0 & 0.1 & 0.05 & 0.0 & 0.01 & STIS/E140M & 4 \\
GJ 412 B & 168.51 & 63.08 & 4.91 & 17.86 & 0.0 & 0.1 & 0.05 & 0.0 & 0.01 & STIS/E140M & 4 \\
AD Leo & 216.46 & 54.58 & 4.97 & 18.48 & 0.001 & 0.1 & 0.19 & 0.0004 & 0.04 & STIS/E140M & 12 \\
GJ 832 & 349.17 & -46.35 & 4.97 & 18.2 & 0.03 & 0.3 & 0.1 & 0.007 & 0.07 & STIS/G140M & 13 \\
40 Eri & 200.75 & -38.05 & 5.01 & 17.85 & 0.1 & 0.1 & 0.05 & 0.01 & 0.01 & GHRS/ECH-A & 9 \\
40 Eri C & 200.77 & -38.03 & 5.01 & 17.79 & 0.0 & 0.1 & 0.04 & 0.0 & 0.009 & STIS/E140M & 4 \\
EV Lac & 100.61 & -13.07 & 5.05 & 17.97 & 0.007 & 0.1 & 0.06 & 0.001 & 0.01 & STIS/E140M & 12 \\
70 Oph A & 29.89 & 11.37 & 5.11 & 18.06 & 0.002 & 0.1 & 0.07 & 0.0003 & 0.02 & STIS/E140M & 12 \\
$\alpha$ Aql & 47.74 & -8.91 & 5.13 & 18.3 & 0.36 & 0.36 & 0.12 & 0.1 & 0.1 & GHRS/G140M & 14 \\
GJ 3323 & 206.43 & -27.47 & 5.37 & 17.72 & 0.02 & 0.1 & 0.03 & 0.001 & 0.007 & STIS/E140M & 15 \\
LSR J1835+3259 & 61.65 & 17.44 & 5.69 & 18.0 & 0.0 & 0.3 & 0.06 & 0.0 & 0.04 & STIS/G140M & 16 \\
GJ 205 & 206.93 & -19.45 & 5.7 & 17.76 & 0.02 & 0.1 & 0.03 & 0.002 & 0.008 & STIS/E140M & 8 \\
GJ 754 & 352.36 & -23.9 & 5.91 & 18.27 & 0.008 & 0.1 & 0.1 & 0.002 & 0.02 & STIS/E140M & 17 \\
GJ 588 & 332.67 & 12.13 & 5.92 & 18.12 & 0.01 & 0.1 & 0.07 & 0.002 & 0.02 & STIS/E140M & 8 \\
36 Oph A & 358.28 & 6.88 & 5.95 & 17.85 & 0.15 & 0.15 & 0.04 & 0.01 & 0.01 & STIS/E140H & 18 \\
YZ CMi & 215.86 & 13.46 & 5.99 & 18.02 & 0.02 & 0.1 & 0.06 & 0.003 & 0.01 & STIS/E140M & 8 \\
HD 191408 & 5.23 & -30.92 & 6.01 & 18.28 & 0.02 & 0.1 & 0.1 & 0.005 & 0.02 & STIS/E140M & 3 \\
82 Eri & 250.75 & -56.08 & 6.04 & 18.33 & 0.03 & 0.1 & 0.11 & 0.008 & 0.03 & STIS/E140M & 3 \\
GJ 780 & 329.77 & -32.42 & 6.1 & 18.01 & 0.003 & 0.1 & 0.05 & 0.0004 & 0.01 & STIS/E140H & 17 \\
GJ 644B & 11.04 & 21.14 & 6.2 & 18.4 & 0.01 & 0.1 & 0.13 & 0.003 & 0.03 & STIS/E140H & 8 \\
GJ 581 & 354.08 & 40.02 & 6.3 & 18.01 & 0.14 & 0.3 & 0.05 & 0.02 & 0.04 & STIS/G140M & 13 \\
GJ338AB & 164.94 & 42.67 & 6.33 & 17.97 & 0.01 & 0.1 & 0.05 & 0.001 & 0.01 & STIS/E140M & 8 \\
$\xi$ Boo A & 23.09 & 61.36 & 6.75 & 17.92 & 0.007 & 0.1 & 0.04 & 0.0006 & 0.009 & STIS/E140M & 12 \\
LTT 1445A & 200.49 & -58.07 & 6.86 & 17.8 & 0.1 & 0.3 & 0.03 & 0.007 & 0.02 & STIS/G140M & 19 \\
GJ 667C & 351.84 & 1.41 & 7.24 & 17.98 & 0.15 & 0.3 & 0.04 & 0.01 & 0.03 & STIS/G140M & 13 \\
GJ 486 & 300.05 & 72.6 & 8.08 & 17.8 & 0.1 & 0.3 & 0.03 & 0.006 & 0.02 & STIS/G140M & 19 \\
GJ 686 & 42.24 & 24.3 & 8.16 & 18.28 & 0.02 & 0.1 & 0.08 & 0.003 & 0.02 & STIS/E140M & 17 \\
61 Vir & 311.86 & 44.09 & 8.53 & 17.91 & 0.04 & 0.1 & 0.03 & 0.003 & 0.007 & STIS/E140M & 12 \\
$\chi^1$ Ori & 188.46 & -2.73 & 8.66 & 17.93 & 0.003 & 0.1 & 0.03 & 0.0002 & 0.007 & STIS/E140M & 12 \\
HD 192310 & 15.62 & -29.4 & 8.81 & 18.24 & 0.02 & 0.1 & 0.06 & 0.003 & 0.01 & STIS/E140M & 20 \\
GJ 849 & 55.89 & -45.4 & 8.81 & 17.83 & 0.07 & 0.3 & 0.02 & 0.004 & 0.02 & STIS/G140M & 6 \\
HD 32147 & 205.09 & -27.18 & 8.84 & 17.84 & 0.14 & 0.14 & 0.03 & 0.008 & 0.008 & STIS/E140M & 15 \\
delta Eri & 198.09 & -46.0 & 9.09 & 17.88 & 0.007 & 0.1 & 0.03 & 0.0004 & 0.006 & STIS/E140M & 12 \\
$\kappa^1$ Cet & 178.22 & -43.07 & 9.28 & 17.89 & 0.003 & 0.1 & 0.03 & 0.0002 & 0.006 & STIS/E140M & 12 \\
GJ 367 & 272.24 & 5.66 & 9.42 & 17.78 & 0.09 & 0.3 & 0.02 & 0.004 & 0.01 & STIS/G140M & 21 \\
L 678-39 & 253.88 & 22.08 & 9.44 & 18.66 & 0.17 & 0.3 & 0.16 & 0.06 & 0.11 & STIS/G140M & 22 \\
GJ 176 & 180.02 & -17.43 & 9.49 & 17.46 & 0.09 & 0.3 & 0.01 & 0.002 & 0.007 & STIS/G140M & 13 \\
AU Mic & 12.65 & -36.8 & 9.71 & 18.36 & 0.002 & 0.1 & 0.08 & 0.0003 & 0.02 & STIS/E140M & 12 \\
GJ 436 & 210.54 & 74.57 & 9.78 & 18.04 & 0.06 & 0.3 & 0.04 & 0.005 & 0.03 & STIS/G140M & 13 \\
GJ 678.1 A & 28.57 & 20.54 & 10.12 & 18.58 & 0.0003 & 0.1 & 0.12 & 7e-05 & 0.03 & STIS/E140M & 17 \\
$\beta$ Gem & 192.23 & 23.41 & 10.36 & 18.26 & 0.03 & 0.1 & 0.06 & 0.004 & 0.01 & GHRS/ECH-A & 7 \\
EP Eri & 192.07 & -58.25 & 10.36 & 18.05 & 0.1 & 0.3 & 0.04 & 0.008 & 0.02 & GHRS/G160M & 23 \\
GJ 649 & 46.52 & 35.34 & 10.39 & 18.19 & 0.02 & 0.3 & 0.05 & 0.002 & 0.03 & STIS/G140M & 6 \\
GJ 341 & 279.59 & -7.3 & 10.45 & 18.37 & 0.06 & 0.3 & 0.07 & 0.01 & 0.05 & STIS/G140M & 21 \\
GJ 173 & 207.62 & -34.65 & 11.21 & 17.86 & 0.02 & 0.1 & 0.02 & 0.001 & 0.005 & STIS/E140M & 15 \\
HD 85512 & 271.68 & 8.16 & 11.28 & 18.43 & 0.12 & 0.12 & 0.08 & 0.02 & 0.02 & STIS/E140M & 13 \\
L 802-6 & 193.58 & -57.14 & 11.5 & 18.39 & 0.33 & 0.33 & 0.07 & 0.05 & 0.05 & STIS/G140M & 24 \\
$\zeta$ Dor & 266.03 & -36.72 & 11.69 & 18.29 & 0.002 & 0.1 & 0.05 & 0.0002 & 0.01 & STIS/E140M & 12 \\
LHS 2686 & 114.08 & 69.05 & 12.19 & 17.34 & 0.13 & 0.3 & 0.006 & 0.002 & 0.004 & STIS/G140M & 6 \\
HR 1925 & 158.38 & 11.95 & 12.27 & 18.26 & 0.005 & 0.1 & 0.05 & 0.0005 & 0.01 & STIS/E140M & 12 \\
TRAPPIST-1 & 69.71 & -56.64 & 12.47 & 18.4 & 0.1 & 0.3 & 0.07 & 0.02 & 0.05 & STIS/G140M & 25 \\
GJ 1132 & 277.26 & 7.76 & 12.61 & 17.49 & 0.62 & 0.62 & 0.008 & 0.01 & 0.01 & STIS/G140M & 6 \\
HD 40307 & 268.81 & -30.34 & 12.93 & 18.6 & 0.08 & 0.1 & 0.1 & 0.02 & 0.02 & STIS/E140M & 13 \\
Capella & 162.59 & 4.57 & 13.12 & 18.24 & 0.006 & 0.1 & 0.04 & 0.0006 & 0.01 & GHRS/ECH-A & 10 \\
HD 9826 & 132.0 & -20.67 & 13.48 & 18.07 & 0.13 & 0.13 & 0.03 & 0.008 & 0.008 & STIS/E140M & 20 \\
HR 8 & 111.26 & -32.83 & 13.77 & 18.29 & 0.006 & 0.1 & 0.05 & 0.0006 & 0.01 & STIS/E140M & 12 \\
$\pi^1$ UMa & 150.55 & 35.7 & 14.44 & 18.12 & 0.01 & 0.1 & 0.03 & 0.0007 & 0.007 & STIS/E140M & 26 \\
HD 35296 & 187.2 & -10.28 & 14.58 & 17.85 & 0.02 & 0.1 & 0.02 & 0.0007 & 0.004 & STIS/E140M & 27 \\
GJ 1214 & 26.16 & 23.61 & 14.64 & 18.06 & 0.66 & 0.66 & 0.03 & 0.04 & 0.04 & STIS/G140M & 13 \\
GJ 163 & 262.78 & -45.32 & 15.14 & 18.48 & 0.06 & 0.3 & 0.06 & 0.009 & 0.04 & STIS/G140M & 6 \\
$\chi$ Her & 67.7 & 50.32 & 15.9 & 18.25 & 0.003 & 0.1 & 0.04 & 0.0002 & 0.008 & STIS/E140M & 12 \\
GJ676A & 339.1 & -9.54 & 15.98 & 18.24 & 0.02 & 0.3 & 0.04 & 0.002 & 0.02 & STIS/G140M & 6 \\
HR 2225 & 230.85 & -18.52 & 16.7 & 17.87 & 0.01 & 0.1 & 0.01 & 0.0003 & 0.003 & STIS/E140M & 12 \\
$\beta$ Cas & 117.53 & -3.28 & 16.78 & 18.13 & 0.05 & 0.1 & 0.03 & 0.003 & 0.006 & GHRS/ECH-A & 7 \\
HR 6748 & 356.04 & -7.33 & 17.11 & 18.13 & 0.02 & 0.1 & 0.03 & 0.001 & 0.006 & STIS/E140M & 12 \\
$\iota$ Hor & 268.81 & -58.32 & 17.36 & 18.81 & 0.05 & 0.1 & 0.12 & 0.01 & 0.03 & STIS/E140M & 28 \\
DX Leo & 201.21 & 46.06 & 18.07 & 17.7 & 0.15 & 0.3 & 0.009 & 0.003 & 0.006 & GHRS/G160M & 23 \\
HD 206860 & 69.86 & -28.27 & 18.13 & 18.35 & 0.12 & 0.12 & 0.04 & 0.01 & 0.01 & STIS/E140M & 20 \\
LQ Hya & 244.59 & 28.4 & 18.27 & 19.05 & 0.15 & 0.3 & 0.2 & 0.07 & 0.14 & GHRS/G160M & 23 \\
$\pi$ Men & 292.51 & -29.78 & 18.29 & 18.51 & 0.02 & 0.3 & 0.06 & 0.003 & 0.04 & STIS/G140M & 29 \\
V368 Cep & 118.46 & 16.94 & 18.95 & 17.95 & 0.1 & 0.3 & 0.02 & 0.004 & 0.01 & GHRS/G160M & 23 \\
$\alpha$ Tri & 138.64 & -31.41 & 19.48 & 18.33 & 0.07 & 0.1 & 0.04 & 0.006 & 0.008 & GHRS/ECH-A & 7 \\
HD 189733 & 60.96 & -3.92 & 19.78 & 18.48 & 0.08 & 0.3 & 0.05 & 0.009 & 0.03 & STIS/G140M & 30 \\
TOI-260 & 97.45 & -71.22 & 20.21 & 17.8 & 0.14 & 0.3 & 0.01 & 0.003 & 0.007 & STIS/G140M & 21 \\
$\alpha$ Tau & 180.97 & -20.25 & 20.43 & 18.29 & 0.0 & 0.1 & 0.03 & 0.0 & 0.007 & STIS/E140M & 31 \\
HD 203244 & 324.9 & -38.91 & 20.81 & 18.82 & 0.01 & 0.1 & 0.1 & 0.002 & 0.02 & STIS/E140M & 12 \\
HIP 112312 & 12.51 & -62.11 & 20.87 & 18.28 & 0.02 & 0.1 & 0.03 & 0.001 & 0.007 & STIS/E140M & 32 \\
HD 97658 & 211.16 & 68.3 & 21.56 & 18.45 & 0.12 & 0.12 & 0.04 & 0.01 & 0.01 & STIS/E140M & 13 \\
HD 116956 & 113.7 & 59.54 & 21.65 & 18.19 & 0.009 & 0.1 & 0.02 & 0.0005 & 0.005 & STIS/E140M & 12 \\
HD 59967 & 250.45 & -8.97 & 21.75 & 18.54 & 0.008 & 0.1 & 0.05 & 0.0009 & 0.01 & STIS/E140M & 12 \\
HD 106516 & 288.5 & 51.54 & 22.35 & 18.57 & 0.02 & 0.1 & 0.05 & 0.002 & 0.01 & STIS/E140M & 12 \\
HD 97334 & 184.33 & 67.28 & 22.7 & 17.82 & 0.01 & 0.1 & 0.009 & 0.0002 & 0.002 & STIS/E140M & 12 \\
HD 128987 & 337.54 & 39.22 & 23.73 & 18.11 & 0.02 & 0.1 & 0.02 & 0.0008 & 0.004 & STIS/E140M & 12 \\
$\alpha$ Leo & 226.43 & 48.93 & 24.31 & 18.28 & 0.17 & 0.17 & 0.03 & 0.01 & 0.01 & STIS/E140H & 33 \\
HD 73350 & 232.08 & 19.98 & 24.35 & 18.16 & 0.009 & 0.1 & 0.02 & 0.0004 & 0.004 & STIS/E140M & 12 \\
Ross 451 & 133.18 & 48.43 & 24.74 & 18.41 & 0.38 & 0.38 & 0.03 & 0.03 & 0.03 & STIS/G140M & 24 \\
TOI-134 & 321.95 & -53.56 & 25.18 & 18.71 & 0.08 & 0.3 & 0.07 & 0.01 & 0.05 & STIS/G140M & 21 \\
$\lambda$ And & 109.9 & -14.54 & 25.92 & 18.45 & 0.15 & 0.15 & 0.04 & 0.01 & 0.01 & GHRS/ECH-A & 11 \\
Ross 1044 & 2.21 & 50.8 & 26.71 & 18.86 & 0.09 & 0.3 & 0.09 & 0.02 & 0.06 & STIS/G140M & 34 \\
HIP 117795 & 115.71 & -2.11 & 26.73 & 18.13 & 0.48 & 0.48 & 0.02 & 0.02 & 0.02 & STIS/G140M & 24 \\
TOI-776 & 290.57 & 23.94 & 27.15 & 18.65 & 0.06 & 0.3 & 0.05 & 0.007 & 0.04 & STIS/G140M & 21 \\
HD 64090 & 190.16 & 25.93 & 27.35 & 18.46 & 0.16 & 0.3 & 0.03 & 0.01 & 0.02 & STIS/G140M & 24 \\
TOI-1231 & 281.83 & 4.38 & 27.48 & 18.75 & 0.15 & 0.3 & 0.07 & 0.02 & 0.05 & STIS/G140M & 35 \\
PW And & 114.62 & -31.38 & 28.27 & 18.35 & 0.1 & 0.3 & 0.03 & 0.006 & 0.02 & GHRS/G160M & 23 \\
EG UMa & 135.26 & 63.75 & 28.61 & 17.5 & 0.3 & 0.3 & 0.004 & 0.002 & 0.002 & STIS/G140M & 36 \\
HD 134440 & 344.67 & 34.93 & 29.39 & 18.51 & 0.36 & 0.36 & 0.04 & 0.03 & 0.03 & STIS/G140M & 24 \\
HD 134439 & 344.74 & 34.99 & 29.4 & 18.83 & 0.11 & 0.3 & 0.07 & 0.02 & 0.05 & STIS/G140M & 24 \\
GJ 3470 & 206.24 & 21.76 & 29.4 & 18.07 & 0.09 & 0.3 & 0.01 & 0.003 & 0.009 & STIS/G140M & 37 \\
HR 1099 & 184.91 & -41.57 & 29.43 & 18.13 & 0.04 & 0.1 & 0.01 & 0.001 & 0.003 & STIS/E140M & 38 \\
$\beta$ Cet & 111.33 & -80.68 & 29.53 & 18.36 & 0.1 & 0.1 & 0.03 & 0.006 & 0.006 & GHRS/ECH-A & 38 \\
GJ 9827 & 81.34 & -57.18 & 29.65 & 18.22 & 0.09 & 0.3 & 0.02 & 0.004 & 0.01 & STIS/G140M & 39 \\
HD 184499 & 67.03 & 6.54 & 31.83 & 18.85 & 0.08 & 0.1 & 0.07 & 0.01 & 0.02 & STIS/E140M & 35 \\
$\eta$ UMa & 100.7 & 65.32 & 31.87 & 17.92 & 0.04 & 0.1 & 0.009 & 0.0008 & 0.002 & GHRS/ECH-A & 14 \\
DK UMa & 142.55 & 38.93 & 32.1 & 18.05 & 0.002 & 0.1 & 0.01 & 5e-05 & 0.003 & STIS/E140M & 12 \\
EK Dra & 105.52 & 49.04 & 34.4 & 18.0 & 0.04 & 0.1 & 0.009 & 0.0009 & 0.002 & STIS/E140M & 40 \\
$\mu$ Vel & 283.03 & 8.57 & 35.92 & 18.47 & 0.001 & 0.1 & 0.03 & 6e-05 & 0.006 & STIS/E140M & 12 \\
Kepler 444 & 73.36 & 12.89 & 36.55 & 18.65 & 0.25 & 0.3 & 0.04 & 0.02 & 0.03 & STIS/G140M & 41 \\
$\sigma$ Gem & 191.19 & 23.27 & 36.87 & 18.2 & 0.07 & 0.1 & 0.01 & 0.002 & 0.003 & GHRS/ECH-A & 7 \\
HAT-P-11 & 81.83 & 10.77 & 37.84 & 18.65 & 0.0 & 0.3 & 0.04 & 0.0 & 0.03 & STIS/G140M & 42 \\
K2-18 & 254.63 & 62.58 & 38.1 & 18.16 & 0.39 & 0.39 & 0.01 & 0.01 & 0.01 & STIS/G140M & 43 \\
HD 19445 & 157.48 & -27.2 & 38.65 & 17.71 & 0.16 & 0.3 & 0.004 & 0.002 & 0.003 & STIS/G140M & 35 \\
DS Tuc A & 312.1 & -46.63 & 44.18 & 18.8 & 0.04 & 0.3 & 0.05 & 0.004 & 0.03 & STIS/G140M & 44 \\
HD 28568 & 180.46 & -21.44 & 44.54 & 18.12 & 0.02 & 0.1 & 0.01 & 0.0004 & 0.002 & STIS/E140M & 12 \\
K2-25 & 178.26 & -25.29 & 44.73 & 18.14 & 0.08 & 0.3 & 0.01 & 0.002 & 0.007 & STIS/G140M & 45 \\
V993 Tau & 180.41 & -22.39 & 46.98 & 18.11 & 0.008 & 0.1 & 0.009 & 0.0002 & 0.002 & STIS/E140M & 12 \\
HD 3167 & 115.11 & -58.24 & 47.31 & 18.38 & 0.16 & 0.3 & 0.02 & 0.006 & 0.01 & STIS/G140M & 46 \\
HD 28033 & 175.37 & -18.86 & 47.57 & 18.15 & 0.02 & 0.1 & 0.01 & 0.0004 & 0.002 & STIS/E140M & 12 \\
V471 Tau & 172.48 & -27.94 & 47.59 & 18.21 & 0.007 & 0.1 & 0.01 & 0.0002 & 0.003 & STIS/E140M & 12 \\
HD 209458 & 76.75 & -28.53 & 48.15 & 18.37 & 0.02 & 0.1 & 0.02 & 0.0007 & 0.004 & STIS/E140M & 12 \\
Speedy Mic & 6.25 & -38.25 & 51.02 & 18.3 & 0.15 & 0.3 & 0.01 & 0.004 & 0.009 & GHRS/G160M & 23 \\
G191-B2B & 155.95 & 7.1 & 52.49 & 18.18 & 0.2 & 0.2 & 0.009 & 0.004 & 0.004 & STIS/E140H & 47 \\
$\upsilon$ Peg & 98.56 & -35.36 & 53.46 & 18.24 & 0.001 & 0.1 & 0.01 & 2e-05 & 0.002 & STIS/E140M & 12 \\
HD 32008 & 209.55 & -29.41 & 54.05 & 18.07 & 0.04 & 0.1 & 0.007 & 0.0006 & 0.002 & STIS/E140M & 12 \\
HD 194598 & 52.82 & -16.13 & 54.23 & 17.96 & 0.33 & 0.33 & 0.005 & 0.004 & 0.004 & STIS/G140M & 35 \\
$\alpha$ Tuc & 330.22 & -47.96 & 56.39 & 18.79 & 0.0 & 0.1 & 0.04 & 0.0 & 0.008 & STIS/E140M & 31 \\
$\sigma$ Pup & 255.74 & -11.91 & 58.74 & 18.87 & 0.0 & 0.1 & 0.04 & 0.0 & 0.009 & STIS/E140M & 31 \\
K2-136 & 174.77 & -17.35 & 58.89 & 18.26 & 0.4 & 0.4 & 0.01 & 0.009 & 0.009 & STIS/G140M & 46 \\
HZ 43 & 54.11 & 84.16 & 59.68 & 17.93 & 0.06 & 0.1 & 0.005 & 0.0006 & 0.001 & GHRS/ECH-A & 48 \\
$\iota$ Cap & 33.63 & -40.77 & 61.71 & 18.7 & 0.001 & 0.1 & 0.03 & 6e-05 & 0.006 & STIS/E140M & 12 \\
HIP 116454 & 86.46 & -56.97 & 62.62 & 17.92 & 0.09 & 0.3 & 0.004 & 0.0009 & 0.003 & STIS/G140M & 46 \\
TOI-178 & 357.3 & -84.1 & 62.89 & 18.6 & 0.29 & 0.3 & 0.02 & 0.01 & 0.01 & STIS/G140M & 21 \\
WASP-107 & 295.69 & 52.47 & 64.4 & 17.54 & 0.64 & 0.64 & 0.002 & 0.003 & 0.003 & STIS/G140M & 46 \\
GD 153 & 317.26 & 84.75 & 68.53 & 17.95 & 0.03 & 0.1 & 0.004 & 0.0003 & 0.001 & STIS/E140H & 14 \\
GD 246 & 87.26 & -45.11 & 75.17 & 19.11 & 0.05 & 0.1 & 0.06 & 0.006 & 0.01 & STIS/E140M & 49 \\
HD 149026 & 61.3 & 43.28 & 76.22 & 18.69 & 0.2 & 0.3 & 0.02 & 0.01 & 0.01 & STIS/G140M & 22 \\
Feige 24 & 165.97 & -50.27 & 77.71 & 18.47 & 0.03 & 0.1 & 0.01 & 0.0008 & 0.003 & STIS/E140M & 50 \\
HR 3145 & 219.07 & 16.8 & 79.85 & 18.56 & 0.0 & 0.1 & 0.01 & 0.0 & 0.003 & STIS/E140M & 31 \\
$\theta$ CMa & 223.98 & -4.86 & 82.87 & 17.97 & 0.0 & 0.1 & 0.004 & 0.0 & 0.0008 & STIS/E140M & 31 \\
31 Com & 114.95 & 89.58 & 84.41 & 17.88 & 0.06 & 0.1 & 0.003 & 0.0004 & 0.0007 & GHRS/ECH-A & 7 \\
WASP-29 & 343.32 & -72.18 & 87.45 & 18.55 & 0.04 & 0.3 & 0.01 & 0.001 & 0.009 & STIS/G140M & 51 \\
HR 3980 & 228.95 & 47.89 & 90.74 & 19.53 & 0.0 & 0.1 & 0.12 & 0.0 & 0.03 & STIS/E140M & 31 \\
Ross 825 & 79.01 & -9.98 & 98.26 & 18.8 & 0.15 & 0.3 & 0.02 & 0.007 & 0.01 & STIS/G140M & 34 \\
\enddata
\tablecomments{Star names, Galactic coordinates, and distances are from SIMBAD \citep{Simbad}. Column densities \logNHI\ and uncertainties $\sigma_{N(HI)}$ are derived from \Lya\ spectra from HST STIS or GHRS instruments, {as reported in the literature. The resolving powers ($\lambda$/$\Delta \lambda$) of each mode are approximately 10,000 (STIS/G140M), 15,000 (GHRS/G160M), {20,000 (GHRS/G140M),} 45,000 (STIS/E140M), 80,000 (GHRS/ECH-A), 110,000 (STIS/E140H). The adjusted uncertainties ($\sigma_{N(HI)}^{\rm adj.}$) impose a minimum uncertainty depending on the mode's resolving power (0.1 dex for $R\geq$45,000 and 0.3 dex for $R\leq$20,000); the adjusted uncertainties are used for deriving the \NHI\ maps.}  Average volume densities \nhiavg\ and uncertainties $\sigma_{\bar{n}(HI)}$ {and $\sigma_{\bar{n}(HI)}^{\rm adj.}$} are also listed.  The reference column lists sources of the column densities and uncertainties: [1] \citealt{Wood2001},
[2] \citealt{Linsky1996},
[3] \citealt{Youngblood2022},
[4] C. Johns-Krull et al. in preparation,
[5] \citealt{Hebrard1999},
[6] \citealt{Wilson2025},
[7] \citealt{Dring1997},
[8] \citealt{Wood2021},
[9] \citealt{Wood1998},
[10] \citealt{Linsky1995a},
[11] \citealt{Wood1996a},
[12] \citealt{Wood2005},
[13] \citealt{Youngblood2016},
[14] \citealt{Redfield2004},
[15] \citealt{Vannier2025},
[16] \citealt{Saur2018},
[17] \citealt{Zachary2018},
[18] \citealt{Wood2000b},
[19] \citealt{Diamond-Lowe2024},
[20] \citealt{Edelman2020},
[21] D. J. Wilson et al. in preparation,
[22] \citealt{Behr2023},
[23] \citealt{Wood2000a},
[24] S. Peacock et al. accepted,
[25] \citealt{Wilson2021},
[26] \citealt{Wood2014},
[27] \citealt{Wood2014ApJ...780..108W},
[28] \citealt{Amazo-Gomez2023},
[29] \citealt{GarciaMunoz2020},
[30] \citealt{Bourrier2020},
[31] \citealt{Wood2024},
[32] \citealt{Youngblood2021a},
[33] \citealt{Gry2017},
[34] \citealt{Schneider2019},
[35] This work,
[36] \citealt{Wilson2022},
[37] \citealt{Bourrier2018_GJ3470b},
[38] \citealt{Piskunov1997},
[39] \citealt{Carleo2021},
[40] \citealt{Shoda2024},
[41] \citealt{Bourrier2017},
[42] \citealt{BenJaffel2022},
[43] \citealt{dosSantos2020},
[44] dos Santos et al. in preparation,
[45] \citealt{Rockcliffe2021},
[46] V. Bourrier private communication,
[47] \citealt{Lemoine2002},
[48] \citealt{Kruk2002},
[49] \citealt{Oliveira2003},
[50] \citealt{Vennes2000},
[51] \citealt{dosSantos2021}. }
\end{deluxetable*}

\begin{figure}[htbp]
    \centering
    \subfigure[K2-136]{
        \includegraphics[clip, trim=0cm 0cm 2.5cm 0cm,width=0.45\textwidth]{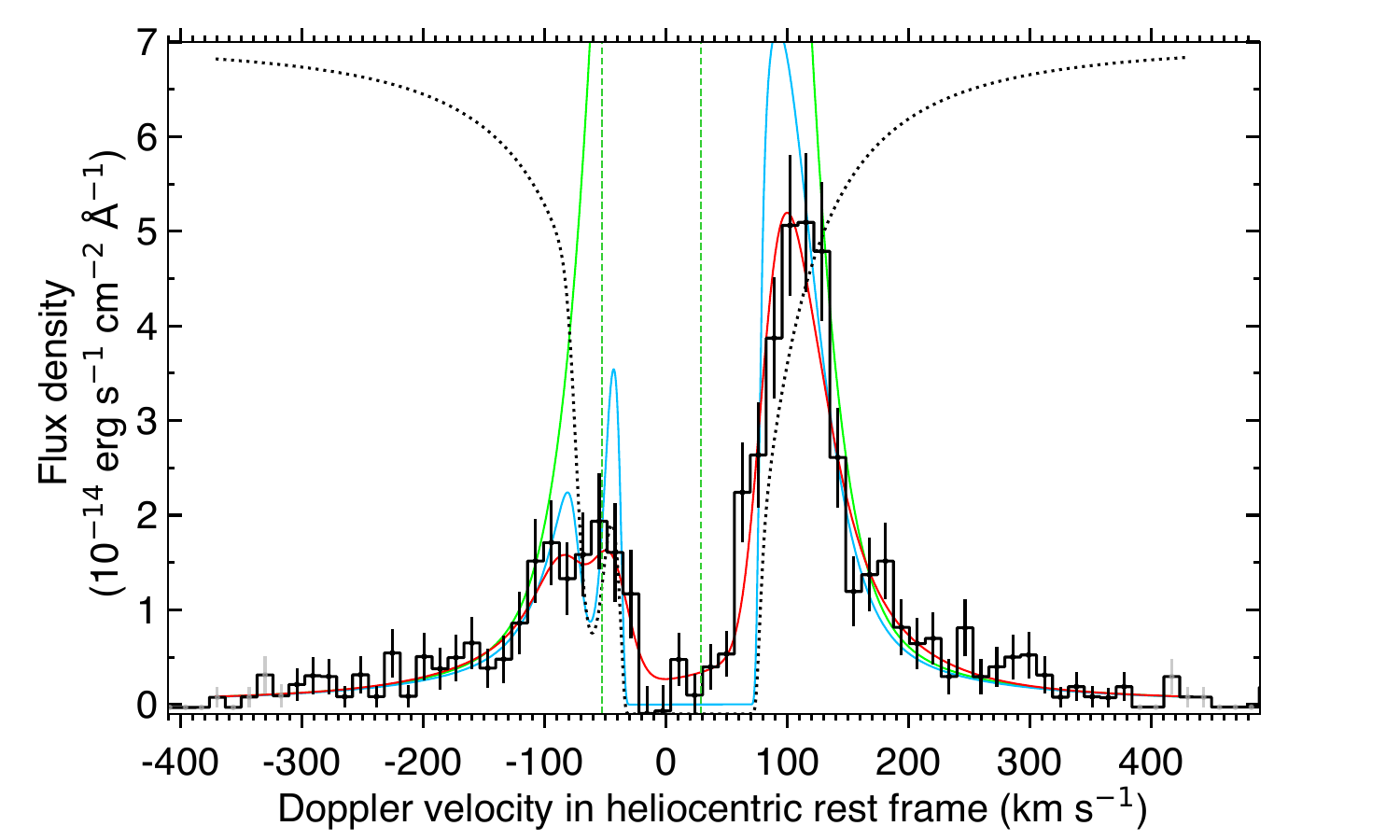}
        \label{fig:subfig1}
    }
    \hfill
    \subfigure[WASP-107]{
        \includegraphics[clip, trim=0cm 0cm 2.5cm 0cm,width=0.45\textwidth]{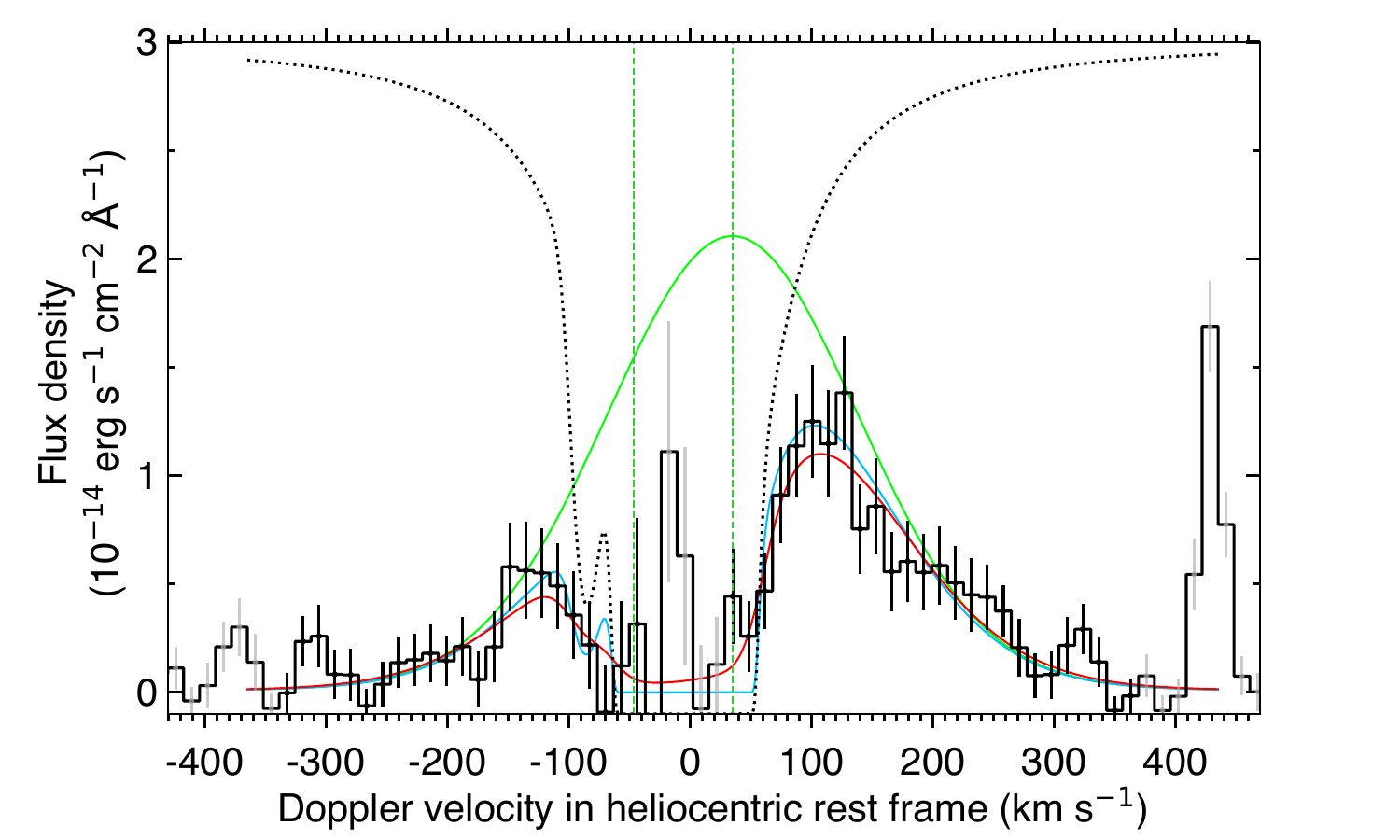}
        \label{fig:subfig2}
    }
    
    \vspace{0.4cm} 
    
    \subfigure[HIP 116454]{
        \includegraphics[clip, trim=0cm 0cm 2.5cm 0cm,width=0.45\textwidth]{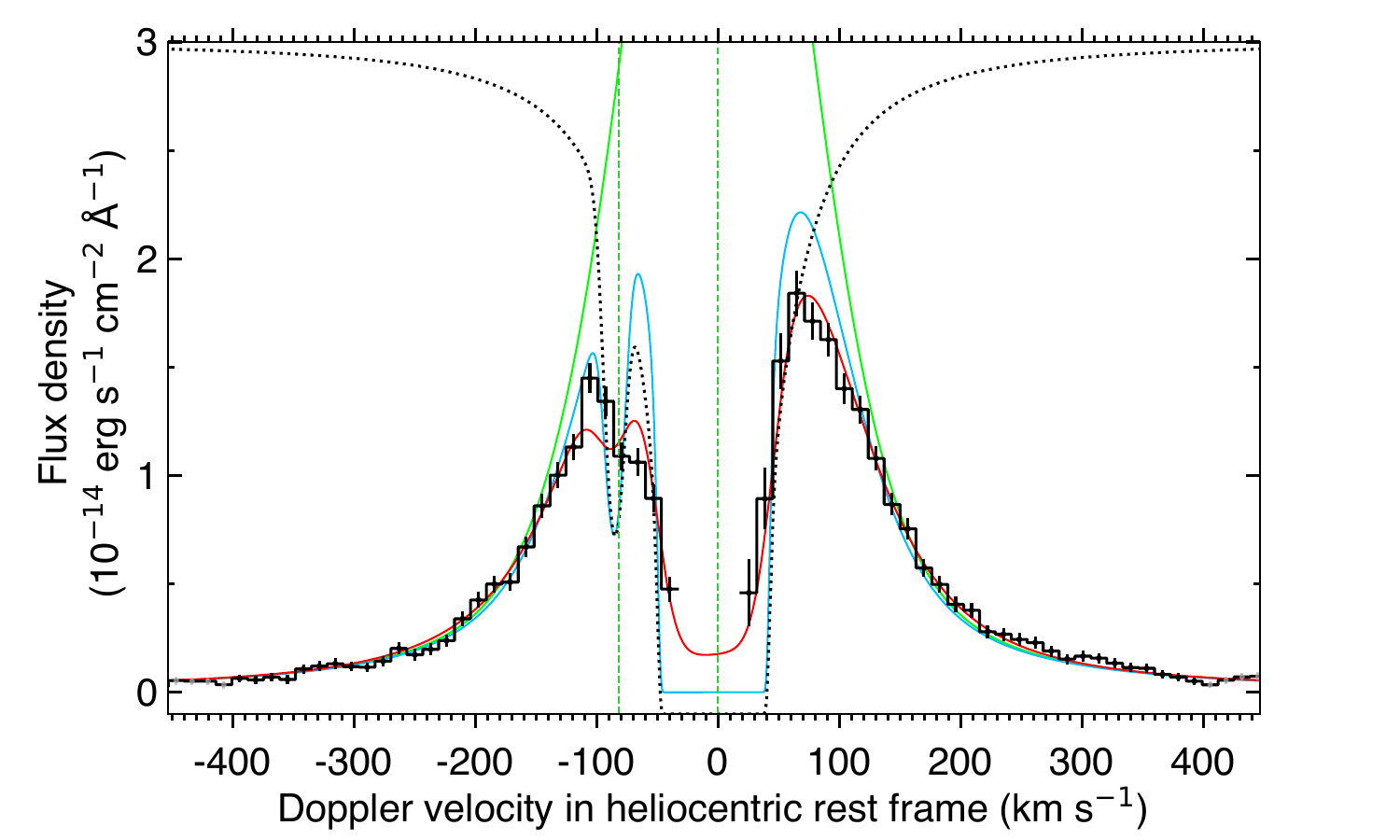}
        \label{fig:subfig3}
    }
    \hfill
    \subfigure[HD 3167]{
        \includegraphics[clip, trim=0cm 0cm 2.5cm 0cm,width=0.45\textwidth]{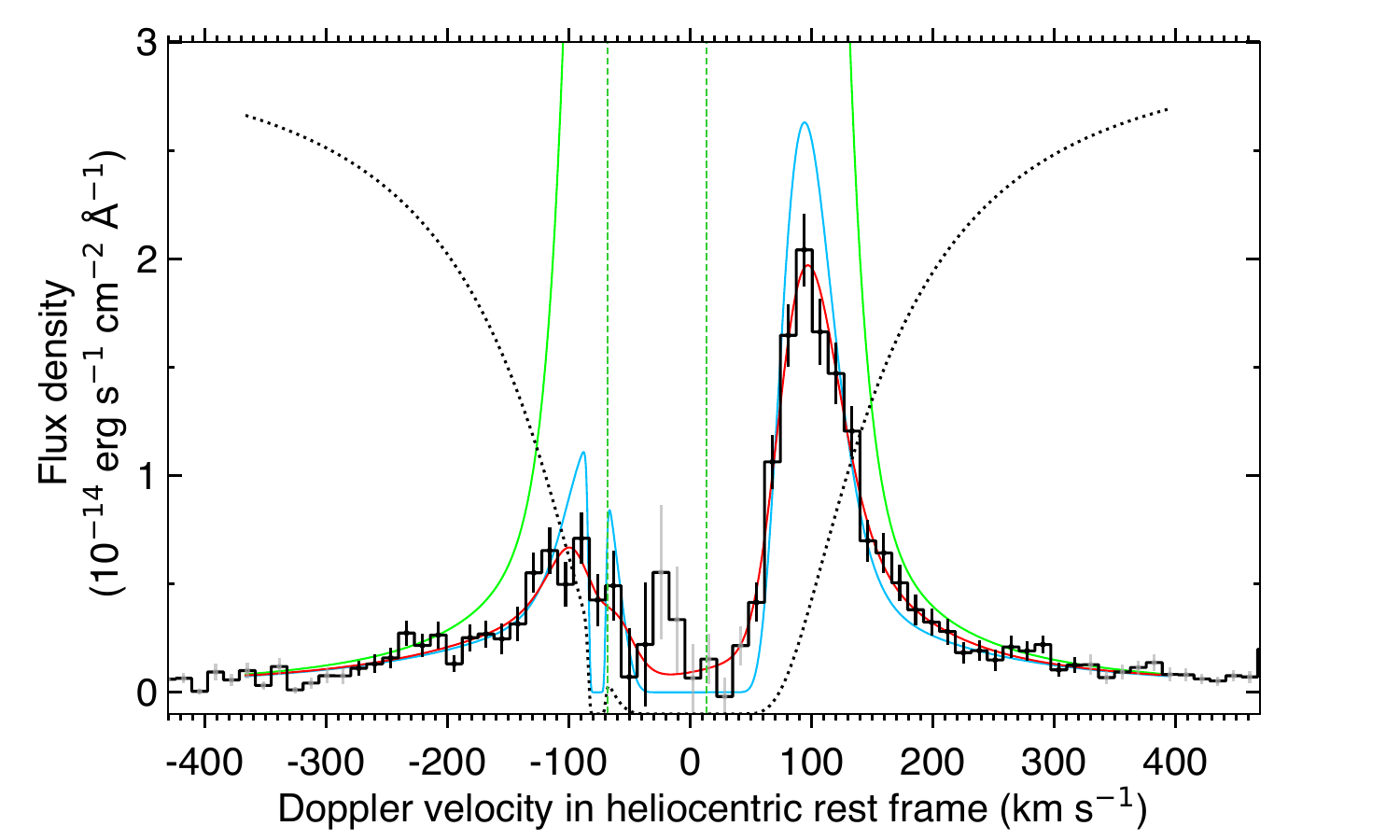}
        \label{fig:subfig4}
    }
    
    \caption{Reconstructions following the methodology of \cite{Bourrier2017, Bourrier2018_GJ3470b, Bourrier2020} are shown for four targets included in this work. The black histograms with error bars show the STIS/G140M data, and the regions marked by the dashed green vertical lines were excluded from the fit due to airglow contamination. The solid green lines show the reconstructed intrinsic stellar profile and the dotted black lines show the transmission of the ISM where the top of the axis corresponds to 100\% transmission and the bottom to 0\% transmission. The solid blue lines show the stellar profile after absorption by the ISM, and the red lines show its convolution with the instrument line spread function. }
    \label{fig:VB_LDS_fits}
\end{figure}

\begin{figure}
    \centering
    \includegraphics[clip, trim=1.5cm 1.5cm 1.5cm 1.5cm, width=\linewidth,angle=180]{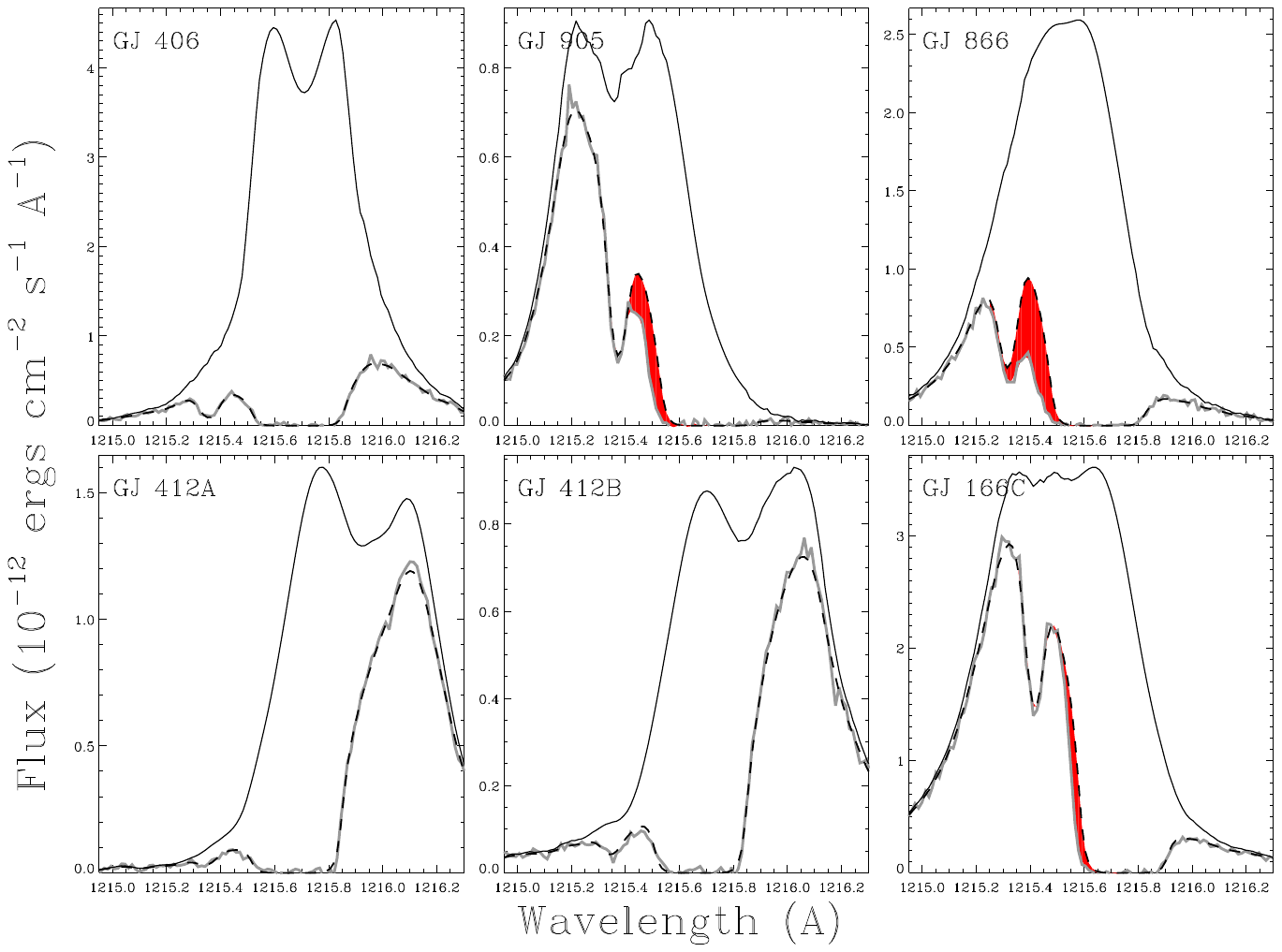}
    \caption{Reconstructions following the methodology of \cite{Wood2005} are shown for six targets included in this work. The light gray lines show the STIS/E140M data, the black dashed lines show the best fit to the data, and the solid black lines show the reconstructed intrinsic stellar profile. Highlighted in red is excess absorption attributed to an astrosphere. These fits will be presented in more detail in C. Johns-Krull et al. in preparation.}
    \label{fig:Wood_in_prep}
\end{figure}

\begin{figure}
    \centering
    \includegraphics[width=\linewidth]{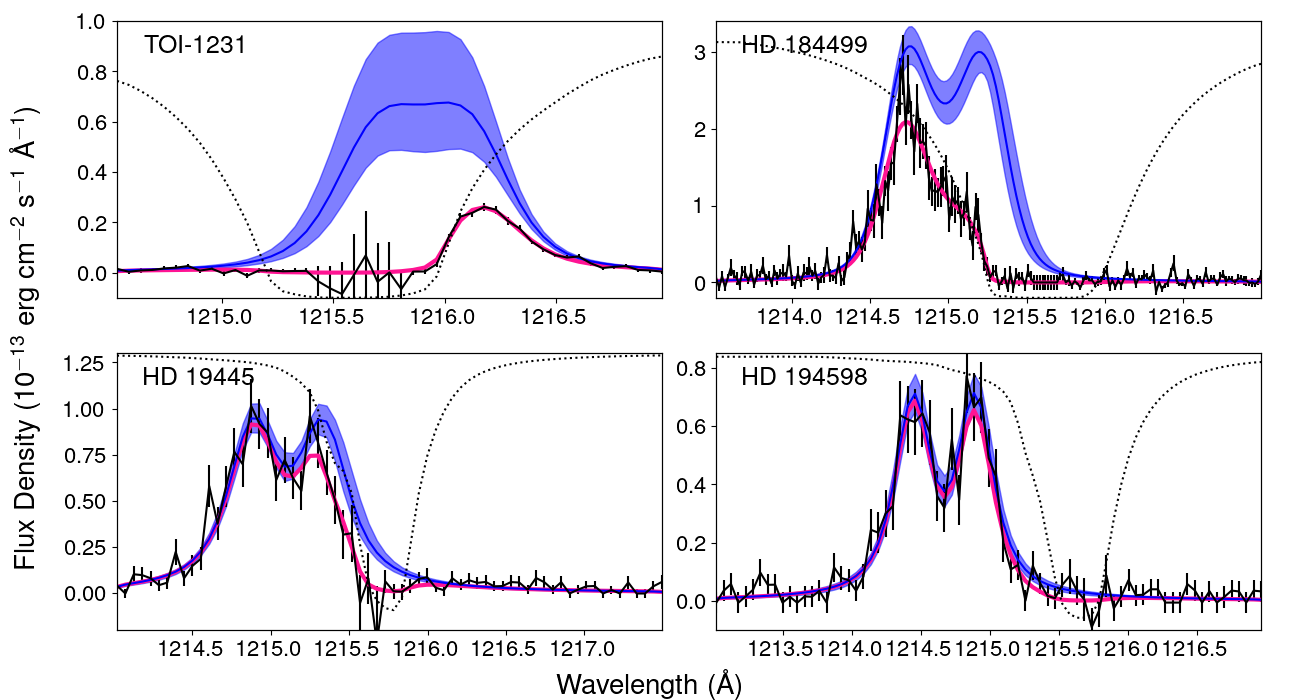}
    \caption{Reconstructions following the methodology of \cite{Youngblood2022} are shown for four targets included in this work. The black lines with error bars show the STIS G140M or E140M data, and the pink lines show the best fit to the data. The blue lines show the reconstructed intrinsic stellar profile and the blue shading shows the uncertainty. The dotted black line shows the transmission of the ISM where the top of the axis corresponds to 100\% transmission and the bottom to 0\% transmission. All curves are shown at the native spectral resolution and binning of the data. The TOI-1231 reconstruction will be presented in more detail in K. Rockcliffe et al. in preparation.}
    \label{fig:AY_fits}
\end{figure}

{Our list of \NHI\ measurements includes 14 from new analyses being performed in parallel with this study, most with involvement from coauthors of this work. Measurements from seven additional systems are incorporated in our analysis, though their reconstructions are reserved for forthcoming dedicated publications (L. dos Santos et al. in prep.; D. Wilson et al. in prep.) and are therefore not displayed here. Figure~\ref{fig:VB_LDS_fits} shows the reconstructions for K2-136, WASP-107, HIP 116454, and HD 3167, which were generated following the methodology detailed in \cite{Bourrier2017, Bourrier2018_GJ3470b, Bourrier2020}, which also provide three of the measurements referenced in this work. Figure~\ref{fig:Wood_in_prep} shows Wolf 359 (GJ 406), GJ 905, EZ Aqr (GJ 866), GJ 412 A, GJ 412 B, and 40 Eri C (GJ 166 C), which were reconstructed following the methodology of \cite{Wood2005} and other works by B. E. Wood, which also provide many of the measurements utilized throughout this study. C. Johns-Krull et al. (in preparation) will present the reconstructions of these six stars in more detail. Figure~\ref{fig:AY_fits} displays the reconstructions of TOI-1231, HD 184499, HD 19445, and HD 194598, which were generated following the methodology of \cite{Youngblood2016, Youngblood2021, Youngblood2022}. These references also provide the basis for other column density measurements cited herein. K. Rockcliffe et al. (in preparation) will present the reconstruction of TOI-1231 in more detail.}

\end{document}